\documentclass[floatfix,preprint,aps,prd,showpacs,amsmath,amssymb]{revtex4}
\usepackage{graphicx}
\usepackage{epsfig}
\usepackage{psfrag}
\newcommand{\be}{\begin{equation}}
\newcommand{\ee}{\end{equation}}

\newcommand{\rd}{\,{\rm d}}
\newcommand{\avec}{\mbox{\boldmath$a$}}

\newcommand{\fvec}{\mbox{\boldmath$f$}}

\newcommand{\dvec}{\mbox{\boldmath$d$}}
\newcommand{\rvec}{\mbox{\boldmath$r$}}
\newcommand{\tvec}{\mbox{\boldmath$t$}}

\newcommand{\D}{\mbox{\boldmath$D$}}
\renewcommand{\b}{\mbox{\boldmath$b$}}
\newcommand{\bepsilon}{\mbox{\boldmath$\epsilon$}}
\begin{document}
%%%%%%%%%%%%%%%%%%%%%%%%%%%%%%%%%%%%%%%%%%%%%%%%%%%%%%%%%%%%%%%%%%%%%
%%%%%%%%%%%%%%%%%%%%%         Title       %%%%%%%%%%%%%%%%%%%%%%%%%%%
%%%%%%%%%%%%%%%%%%%%%%%%%%%%%%%%%%%%%%%%%%%%%%%%%%%%%%%%%%%%%%%%%%%%%
%
\title{Bayesian modeling of source confusion in LISA data}
%
%%%%%%%%%%%%%%%%%%%%%%%%%%%%%%%%%%%%%%%%%%%%%%%%%%%%%%%%%%%%%%%%%%%%%
%%%%%%%%%%%%%%%%%%%%     Authors & Addresses  %%%%%%%%%%%%%%%%%%%%%%%
%%%%%%%%%%%%%%%%%%%%%%%%%%%%%%%%%%%%%%%%%%%%%%%%%%%%%%%%%%%%%%%%%%%%%
%
\author{Richard Umst\"atter$^1$\footnote{richard@stat.auckland.ac.nz},
Nelson Christensen$^2$\footnote{nchriste@carleton.edu}, Martin
Hendry$^3$\footnote{martin@astro.gla.ac.uk}, Renate
Meyer$^1$\footnote{meyer@stat.auckland.ac.nz}, Vimal
Simha$^3$\footnote{vimal\_simha@hotmail.com}, John
Veitch$^3$\footnote{jveitch@astro.gla.ac.uk}, Sarah
Vigeland$^2$\footnote{vigelans@carleton.edu} and Graham
Woan$^3$\footnote{graham@astro.gla.ac.uk} }
\affiliation{$^1$Department of Statistics, University of Auckland,
Auckland, New Zealand\\
$^2$Physics and Astronomy, Carleton College,
Northfield, MN 55057, USA\\
$^3$Department of Physics and Astronomy, University of Glasgow,
Glasgow G12\,8QQ, UK}
\date{\today}
%
%%%%%%%%%%%%%%%%%%%%%%%%%%%%%%%%%%%%%%%%%%%%%%%%%%%%%%%%%%%%%%%%%%%%%%%
%%%%%%%%%%%%%%%%%%%%%           Abstract         %%%%%%%%%%%%%%%%%%%%%%
%%%%%%%%%%%%%%%%%%%%%%%%%%%%%%%%%%%%%%%%%%%%%%%%%%%%%%%%%%%%%%%%%%%%%%%
\begin{abstract}
One of the greatest data analysis challenges for the Laser
Interferometer Space Antenna (LISA) is the need to account for a
large number of gravitational wave signals from compact binary
systems expected to be present in the data. We introduce the basis
of a Bayesian method that we believe can address this challenge,
and demonstrate its effectiveness on a simplified problem
involving one hundred synthetic sinusoidal signals in noise. We
use a reversible jump Markov chain Monte Carlo technique to infer
simultaneously the number of signals present, the parameters of
each identified signal, and the noise level. Our approach
therefore tackles the detection and parameter estimation problems
simultaneously, without the need to evaluate formal model
selection criteria, such as the Akaike Information Criterion or
explicit Bayes factors. The method does not require a stopping
criterion to determine the number of signals, and produces results
which compare very favorably with classical spectral techniques.
\end{abstract}
\pacs{04.80.Nn, 02.70.Lq, 06.20.Dq}
%
%%%%%%%%%%%%%%%%%%%%%%%%%%%%%%%%%%%%%%%%%%%%%%%%%%%%%%%%%%%
%!!!!!!!!!!!!!!!!!!!!  Introduction  !!!!!!!!!!!!!!!!!!!!!%
%%%%%%%%%%%%%%%%%%%%%%%%%%%%%%%%%%%%%%%%%%%%%%%%%%%%%%%%%%%
%
\maketitle
\section{Introduction}
The Laser Interferometer Space Antenna (LISA) is designed to
detect gravitational radiation from astrophysical sources in the
$10^{-2}$\,mHz to $100$\,mHz band \cite{LISA}.  The sensitivity of
LISA is such that very many such sources should be detectable.
Paradoxically, this sensitivity will also make signal
identification somewhat problematic: parts of the band will likely
be swamped with tens of thousands of signals. One of the most
abundant classes of source will be close-by white dwarf binaries,
producing signals from 0.1\,mHz to 3\,mHz. There will be source
confusion below 1\,mHz and resolvable sources above 5\,mHz; the
1\,mHz to 5\,mHz band for LISA therefore presents a tremendous
data analysis challenge, potentially containing up to $10^5$
sources \cite{cutler,bena,cornish}. LISA's ability to detect and
characterize other astrophysical sources will be greatly helped if
the thousands of background signals from binary systems can be
identified. For a detailed look at the population of binary
systems that produce signals in LISA's operating band, and how
they affect LISA's performance, we direct the reader to Barack and
Cutler \cite{cutler} and to Nelemans et al.\ \cite{nele01}.

We approach this problem from a new direction. Markov chain Monte
Carlo (MCMC) techniques have been demonstrated to be especially
suited to parameter estimation problems involving numerous
parameters \cite{gilks96}. We have previously used the
Metropolis-Hastings (MH) algorithm \cite{metr53,hastings70} in
other gravitational radiation problems, such as estimating
astrophysical parameters for gravitational wave signals from
coalescing compact binary systems \cite{insp} or pulsars
\cite{pul1,pul2}. It is our belief that MCMC methods could provide
an effective means for identifying and characterizing the
thousands of background binary signals to be found within the LISA
data.

The method that we present in this paper is not a \emph{source
subtraction} approach \cite{cornish2}, but one that identifies and
characterizes binary produced periodic signals in the data.
Signals that are sufficiently large in amplitude will have their
parameters estimated. Sources that are weak will contribute to the
noise; our method also produces an estimate of the overall level
of the noise. We show that the noise level estimate from our
method depends on the inherent detector noise level, and also the
presence of unidentified signals. MCMC methods are robust and
dynamic, and we believe that ultimately it will be possible to use
them with LISA data to estimate simultaneously the parameters
associated with a wide range of source types occurring in the
presence of many thousands of white dwarf binaries.

In this paper we present the results of a simulation study,
comprising a data stream of $m$ sinusoidal signals embedded in
gaussian noise. Although our simulation study is simple, it does
highlight a number of issues relevant to the real LISA data analysis
problem. For example, the number of signals present in the data,
$m$, will be unknown.

Our Bayesian approach does not need to fit each model with $m$
signals, for $m=1,\ldots M$, and then select the best fitting
model via the evaluation of Bayes factors. The evaluation of Bayes
factors \cite{Kass95,Han01} requires computation of the marginal
likelihoods and thus  marginalizations over the parameter vectors
of each model. This is a formidable computational problem when the
dimension of the parameter space is large. A shortcut to the
calculation of Bayes factors, the harmonic mean of the likelihood
values \cite{Newton94}, is known to be unstable because the
inverse likelihood does not possess a finite variance. Other large
sample approximations to the Bayes factors such as the Bayesian
Information Criterion (BIC), also referred to as {\em Schwarz
Criterion}, and the related penalized likelihood ratio model
choice criterion, Akaike Information Criterion (AIC), have been
shown to be inconsistent when the dimension of the parameter space
goes to infinity \cite{Berger03}. The newly developed deviance
information criterion (DIC) \cite{Linde02} is known to be
controversial in mixture models. While MCMC algorithms like the
Gibbs sampler and MH algorithm yield posterior distributions of
the parameters, they do not provide marginal likelihoods. An
indirect method of estimating marginal likelihoods from Gibbs
sampling output has been developed \cite{Chib95} and extended to
output from the MH algorithm \cite{Chib01}. These, however, are
impractical when the number of candidate models is very large, as
is the case for LISA data. Therefore, we use another strategy, the
reversible jump algorithm \cite{Green95}, that samples over the
model {\em and} parameter space in order to estimate posterior
model probabilities/marginal likelihoods. We consider the number
of sinusoids as an additional parameter and determine its marginal
posterior distribution. Its modal value will give us the model
with the most probable number of sinusoids. As illustrated in
Section II, a Bayesian analysis naturally encompasses Occam's
Razor \cite{Jaynes03} and a preference for a simpler (smaller $m$)
model. We thus address and  solve the detection and estimation of
sinusoids simultaneously. In addition, our MCMC method is better
than a classical periodogram at resolving signals that are very
close in frequency, and we provide a detailed discussion of how to
identify these signals. Finally, our method infers the noise level
in the data together with the parameters of each of the $m$
sinusoids.

The problem of identifying an unknown number of sinusoids is neither
new nor simple \cite{Rich,Andrieu99}. Whereas previous studies have
looked for a handful of unknown signals, here we show results for
100 signals. Another benefit of using
MCMC methods is that computation time scales roughly linearly as the
number of parameters increases, and does not show an exponential
increase in time \cite{gilks96}. In the future we will make the
character of the signals more realistic, taking into account the
orbit of the LISA satellites and the nature of the inspiral of
binaries. With our study of sinusoids we hope to inform LISA data
analysis researchers of another possible avenue for characterizing a
large number of background signals.

The rest of the paper is organized as follows: Section
\ref{model_occam} illustrates that Bayesian inference
automatically implements the principle of Occam's Razor. Our
Bayesian method and posterior computational algorithms are
described in Section \ref{TMHA}. In Section~\ref{simulations} we
present the results of this study using synthesized data. We
believe that this method offers great hope for signal extraction
in real LISA data, and this point is discussed in
Section~\ref{disc}.

\section{Occam's Razor}
\label{model_occam} The notion that Bayesian inference naturally
implements a quantitative version of Occam's Razor \cite{Jaynes03}
is well known, but this property is sufficiently central to our
approach to LISA analysis to warrant a very simple demonstration:

We imagine two physical models, $\mathcal{M}_1$ and $\mathcal{M}_2$
constrained by a single datum, $d$. $\mathcal{M}_1$ has one
parameter, $a_1$, to describe the datum. $\mathcal{M}_2$ uses the
sum of two parameters, $s=a_1+a_2$ to describe the same datum.
Clearly, all other things being equal, if the datum is equally
consistent with both models, we would prefer $\mathcal{M}_1$ over
$\mathcal{M}_2$ on the grounds that $\mathcal{M}_1$ is `simpler'.
To see how this naturally occurs within the Bayesian framework we
proceed by considering the ratio of probabilities (the \emph{odds
ratio}) of the two models:
\begin{equation}
\frac{p(\mathcal{M}_1|d) }{p(\mathcal{M}_2|d)}=
 \frac{p(\mathcal{M}_1)}{p(\mathcal{M}_2)}\frac{p(d|\mathcal{M}_1)}
{p(d|\mathcal{M}_2)}.
\end{equation}
If we set the prior odds ratio to unity, reflecting no prior
preference for either model, the right-hand side of the above
equation is simply the ratio of marginal likelihoods (sometimes
called the \emph{evidences}) of the two models.  This ratio is
conventionally known as the \emph{Bayes factor} for the
comparison. We will take the priors for our parameters $a_1$ and
$a_2$ to be each uniform in the range $0\rightarrow R$. Under
$\mathcal{M}_1$ we therefore have $p(a_1|\mathcal{M}_1) = 1/R$.
The likelihood under $\mathcal{M}_1$ is assumed to be
$p(d|a_1,\mathcal{M}_1)=\delta(d-a_1)$, where $\delta$ denotes the
Dirac function, i.e.\ $\delta(x)=1$ if $x=0$ and 0 otherwise. This
yields
\begin{equation}
p(d|\mathcal{M}_1)=\int_0^R\frac{1}{R}\delta(d-a_1)\,\rd a_1=
\begin{cases}
1/R & 0<d<R\\
0 & \text{otherwise}.
\end{cases}
\end{equation}
Under $\mathcal{M}_2$ with parameter  $s=a_1+a_2$, the prior
probability density function (pdf) for $s$ is the convolution of the
two uniform prior pdfs for $a_1$ and $a_2$ and therefore has the
`triangle' form
\begin{equation}
p(s|\mathcal{M}_2) =
\begin{cases}
s/R^2 & 0<s<R\\
2/R-s/R^2  & R<s<2R,
\end{cases}
\end{equation}
so the evidence for $\mathcal{M}_2$ is
\begin{equation}
p(d|\mathcal{M}_2) = \int_0^{2R}p(s|\mathcal{M}_2)\delta(d-s)\,\rd
s =
\begin{cases}
d/R^2 & 0<d<R\\
2/R-d/R^2  & R<d<2R\\
0     & \text{otherwise},
\end{cases}
\end{equation}
\begin{figure}[hbt]
\centerline{\includegraphics[width=9cm]{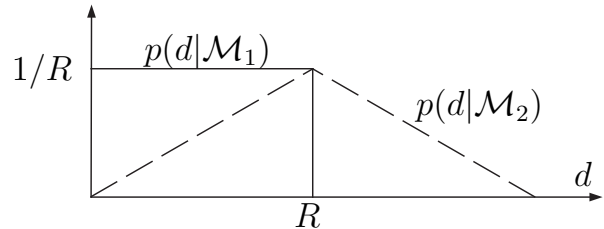} } \caption{The
evidences for models $\mathcal{M}_1$ (solid line, one parameter
fit) and $\mathcal{M}_2$ (dashed line, two parameter fit) as a
function of the datum $d$. Note that the evidence ratio always
favors the simpler model $\mathcal{M}_1$ when the datum is equally
consistent with either.} \label{fig_1}
\end{figure}
(see Fig.~\ref{fig_1}). If the datum lies in the range $0<d<R$, the
Bayes factor is $(1/R)/(d/R^2)=R/d$, and $\mathcal{M}_1$ is always
favoured over $\mathcal{M}_2$.  If $d=R$ neither model is preferred,
and if $R<d<2R$ then $\mathcal{M}_2$ (as the only viable model) is a
relative certainty.  We now see why $\mathcal{M}_1$ is more probable than
$\mathcal{M}_2$ when the datum is equally consistent with both (i.e.,
$d<R$): $\mathcal{M}_2$ has more flexibility than is necessary to
explain the datum and so penalizes itself by spreading its evidence
more thinly.  We can also see that the effect is not simply
dependent on the \emph{number} of parameters. If $\mathcal{M}_2$
depended on just a single parameter but with a uniform prior on this
parameter over $0\rightarrow 2R$, then it too would be discriminated
against (in the case $0<d<R$) for using more parameter space to do
the same job as $\mathcal{M}_1$.

This quantitative  ability to identify the least flexible model
consistent with the data makes Bayesian inference the natural
methodology in problems where the number of parameters is unknown.
Indeed we do not need to consider explicitly Bayes factors at all if
we consider the number of parameters itself to be a parameter, and
determine its marginal posterior probability in the usual way. This
is the approach  we will apply to the the LISA white dwarf binary
problem, as described in the next section.

\section{Extraction of signal parameters}
\label{TMHA} We consider a signal consisting of $m$ superimposed
sinusoidal signals where $m$ is an unknown parameter. Therefore we
confine our attention to a set of models $\{ \mathcal{M}_m: m \in
\{0,\cdots,M\} \}$ with $M$ being the maximum number of sinusoidal
signals we allow. Let  $\dvec=[d_1,\cdots,d_N]$ be a vector of $N$
samples recorded at times $\tvec=[t_1,\cdots,t_N]$. Model
$\mathcal{M}_m$ takes the observed data to comprise a signal,
$s^{(m)}$ plus noise,
$\bepsilon^{(m)}=[\epsilon_1^{(m)},\cdots,\epsilon_N^{(m)}]$:
\begin{equation}\label{model}
d_j=s^{(m)}(t_j,\avec_m)+ \epsilon_j^{(m)}, \quad \mbox{ for }
j=1,\ldots,N
\end{equation}
where the noise terms $\epsilon_j$ are assumed to be i.i.d.\
$N(0,\sigma_m^2)$ random variables. The signal of model
$\mathcal{M}_m$ has the form
\begin{equation}
s^{(m)}(t_j,\avec_m)=\sum_{i=1}^{m}{\left[ A_i^{(m)} \cos(2 \pi
f_i^{(m)} t_j)+B_i^{(m)} \sin(2 \pi f_i^{(m)} t_j) \right]},
\label{signal}
\end{equation}
so that each sinusoid component is characterised by one frequency
and two amplitudes. Model $\mathcal{M}_m$ is therefore characterized
by a vector of $3m+1$ unknown parameters which we denote
$\avec_m=[A_1^{(m)},B_1^{(m)},f_1^{(m)},\cdots,
A_m^{(m)},B_m^{(m)},f_m^{(m)},\sigma^2_m]$. The objective is  to
find the model $\mathcal{M}_m$ that best fits the data and to
estimate its parameters. We use a Bayesian approach as in
{\cite{Bretthorst88}}, but instead of fitting each model separately
and selecting the best using Bayes factors, we treat the number $m$
of unknown sinusoids as an additional unknown parameter and
determine the mode of its marginal posterior distribution (together
with the posterior distribution of all the other parameters). The
joint probability that these data $\dvec$ arise from the parameter
vector $\avec_m$ and model $\mathcal{M}_m$ is given by
\begin{equation}
p(\dvec|m,\avec_m) \propto \frac{1}{\sigma^N_m} \exp \left\{ -
\frac{1}{2 \sigma^2_m} \sum_{j=1}^N \left[d_j -
s^{(m)}(t_j,\avec_m)\right]^2 \right\}. \label{LH}
\end{equation}
For simplicity we take the prior distribution of the model dimension
parameter $m$ as uniform over $\{0,\ldots,M\}$. The variances
$\sigma_m^2$ are given  noninformative inverse gamma priors,
discussed in Section~\ref{updating_the_noise}.

Again, for simplicity our calculations use dimensionless frequencies
with a Nyquist frequency of 0.5, and we use uniform priors for the
component frequencies $f_i^{(m)}$ over $[0,0.5]$.

Given $m$, $\sigma_m^2$ and the frequency vector $\fvec^{(m)}$, our
model (\ref{model}) is a linear regression model which can be
written in matrix form as
\begin{equation}
\dvec=\D^{(m)} \b^{(m)} + \bepsilon^{(m)},
\end{equation}
where
$\b^{(m)}=[A_1^{(m)},B_1^{(m)},A_2^{(m)},B_2^{(m)},\ldots,A_m^{(m)},B_m^{(m)}]$
is the vector of $2m$ amplitudes and the $N\times 2m$ matrix $\D$
contains the entry $\cos(2\pi f_j^{(m)} t_i)$ in row $i$ and column
$2j-1$, and $\sin(2\pi f_j^{(m)} t_i)$ in row $i$ and column $2j$
for $i=1,\ldots, N$ and $j=1,\ldots,m$. Thus, an obvious choice for
the prior distribution of the amplitudes would be a $g$-prior
\cite{Zellner86}, a multivariate Normal distribution with mean zero
and covariance matrix equal to $g\times (\D'\D)^{-1}$. This
$g$-prior was used in \cite{Andrieu99} for situations with $m\leq
4$. However, this choice becomes impractical for a large number $m$
of signals since each iteration of the MCMC algorithm would require
the calculation and inversion of the $2m\times 2m$ covariance matrix
of basis functions. Therefore, for our synthesized data we chose
uniform priors for the amplitudes $A_i^{(m)}$, $B_i^{(m)}$ with
ranges $A_i^{(m)} \in [-A_{\rm max},A_{\rm max}]$ and $ B_i^{(m)}
\in[-B_{\rm max},B_{\rm max}]$ and with $A_{\rm max} = B_{\rm max} =
5$.

Note that these priors are not uninformative, and we express the
parameter space in Cartesian coordinates simply because this is
convenient for the implementation of the MCMC algorithms. A model
expressed in polar coordinates with uniform priors on amplitude
and phase would correspond to a different prior distribution for
the transformed amplitudes in Cartesian coordinates. The effect of
these different priors on the frequency estimates is generally
negligible when the data is informative, and is discussed in
\cite{Bretthorst88}. The posterior pdf of the frequency reaches
its maximum at the same value for both prior choices but with
different curvatures at this maximum.

By applying Bayes' theorem, we obtain the posterior pdf for our
model parameters of
\begin{equation}
p(m,\avec_m|\dvec) = \frac{ p(m,\avec_m) p(\dvec|m,\avec_m)
}{p(\dvec)},
 \label{posterior}
\end{equation}
where $p(\dvec)=\sum_{i=0}^M \int p(m,\avec_m) p(\dvec|m,\avec_m)
\,\rd\avec_m$. The direct evaluation of the normalization constant
$p(\dvec)$ is difficult due to the $(3m+1)$-dimensional
integration involved. Moreover, the computation of marginal
posterior pdfs would require subsequent $3m$-dimensional
integration. To overcome this problem, we use sampling-based MCMC
techniques to carry out posterior inference (see \cite{gilks96}
for an introduction and overview.) These only require the
unnormalized posterior $p(m,\avec_m|\dvec) \propto p(m,\avec_m)
p(\dvec|m,\avec_m)$ to sample from Eq.~(\ref{posterior}) and to
estimate the quantities of interest. However, in the present
context the overall model does not have fixed dimension and
classical MH techniques \cite{metr53,hastings70} cannot be used to
propose trans-dimensional moves. We therefore use the Reversible
Jump Markov Chain Monte Carlo (RJMCMC) algorithm for our model
determination \cite{Green95,Green03}. Additionally we use the
delayed rejection (DR) method \cite{Mira98,TierneyMira99} for
transitions within the same model.  This allows better adaptation
of the proposals in different parts of the state space by allowing
the choice of the proposal distribution to depend on the proposed
but rejected state as well as the current state.

\subsection{The RJMCMC for model determination}
To sample from the joint posterior $p(m,\avec_m|\dvec)$ we
construct a Markov chain simulation with state space
$\displaystyle\cup_{m=1}^M \left({m}\times I\!\!R^{3m+1}\right)$
where $m$ is the current number of signals. When a new model is
proposed we attempt a step between state spaces of different
dimensionality. Suppose that at the $n$th iteration of the Markov
chain we are in state $(k,\avec_k)$. If model $\mathcal{M}_{k'}$
with parameter vector $\avec_{k'}'$ is proposed, a reversible move
has to be considered in order to preserve the detailed balance
equations of the Markov chain. Therefore the dimensions of the
models have to be matched by involving a random vector $\rvec$
sampled from a proposal distribution with pdf $q(\rvec)$, say, for
proposing the  new parameters
$\avec_{k'}'=\textrm{t}(\avec_k,\rvec)$ where $\textrm{t}$ is a
suitable deterministic transformation function of the current
state and $\rvec$. Here we focus on transitions that either
decrease or increase models by one signal, i.e.\ $k' \in
\{k-1,k+1\}$. We use equal probabilities $p_{k\mapsto k'}=p_{{k'
\mapsto k}}$ to either move up or down in dimensionality, and
without loss of generality we consider the upward move $k'=k+1$.

If  the transformation $\textrm{t}_{k\mapsto k'}$ from
$(\avec_k,\rvec)$ to $\avec_{k'}'$ and its inverse
$\textrm{t}_{k\mapsto k'}^{-1}=\textrm{t}_{k'\mapsto k}$ are both
differentiable, then reversibility is guaranteed if we define the
acceptance probability for increasing a model by one signal as
\begin{equation}
\alpha_{k \mapsto k'}(\avec_{k'}'|\avec_k)= \min \left\{ 1, \frac{
p(\avec_{k'}',k') p(\dvec|\avec'_{k'},k')p_{k\mapsto k'}
}{p(\avec_k,k) p(\dvec|\avec_k,k) q(\rvec) p_{k'\mapsto k} }
\right\} \left| J_{k\mapsto k'} \right|
\end{equation}
where $|J_{k\mapsto k'}|=\left|\frac{\partial
\textrm{t}(\avec_{k'},\rvec)}{\partial(\avec_k,\rvec)}\right|$ is the
Jacobian determinant of this transformation \cite{Green95}. In this
context, we suggest two types of transformations, `split-and-merge'
and `birth-and-death'.

\subsection*{Split-and-merge transitions}
For a `split' transition we randomly choose one of our signals
with parameter subvector
$\avec_{(i)}=(A_i^{(k)},B_i^{(k)},f_i^{(k)})$ from $\avec_k$. This
signal is chosen by sampling $i$ uniformly from $\{1,\ldots,k\}$.
The proposed parameter vector $\avec'_{k'}$ comprises all the
other $(k-1)$ subvectors of $\avec_k$  and two additional
3-dimensional subvectors, say
$\avec_{(i_1)}'=(A_{i_1}^{(k')},B_{i_1}^{(k')},f_{i_1}^{(k')})$
and $\avec_{(i_2)}'=(
A_{i_2}^{(k')},B_{i_2}^{(k')},f_{i_2}^{(k')})$ each with half the
amplitude of  $\avec_{(i)}$, but same frequency, to replace
$\avec_{(i)}$. A three-dimensional Gaussian random vector (with
mean zero), $\rvec=(r_A,r_B,r_f)$, changes the current state
$\avec_{(i)}$ to the two resulting states $\avec_{(1i)}',
\avec_{(2i)}'$ through a linear transformation
%\begin{equation} \label{eq:transform}
\begin{eqnarray*}
\textrm{t}_{k\mapsto
k'}(\avec_{(i)},\rvec)&=&\left(\begin{array}{c}
\frac{1}{2}A_i^{(k)} + r_A\\
\frac{1}{2}B_i^{(k)} + r_B\\
f_i^{(k)}+r_f\\
\frac{1}{2}A_i^{(k)} - r_A\\
\frac{1}{2}B_i^{(k)} - r_B\\
f_i^{(k)}-r_f
\end{array}\right)
= \left(\begin{array}{c}
A_{i_1}^{(k')}\\ B_{i_1}^{(k')}\\f_{i_1}^{(k')}\\A_{i_2}^{(k')}\\
B_{i_2}^{(k')}\\f_{i_2}^{(k')}
\end{array}\right).
\end{eqnarray*}
The inverse transformation $\textrm{t}^{-1}_{k\mapsto
k'}:=\textrm{t}_{k'\mapsto k}$ accounts for the merger of two
signals and can be written as
%\begin{equation} \label{invtrans}
\begin{eqnarray*}
\textrm{t}_{k'\mapsto k}
(\avec_{(i_1)}',\avec_{(i_2)}')&=&\left(\begin{array}{c}
A_{i_1}^{(k')} + A_{i_2}^{(k')}\\
B_{i_1}^{(k')} + B_{i_2}^{(k')}\\
\frac{1}{2}f_{i_1}^{(k')} + \frac{1}{2}f_{i_2}^{(k')}\\
\frac{1}{2}(A_{i_1}^{(k')}-A_{i_2}^{(k')})\\
\frac{1}{2}(B_{i_1}^{(k')}-B_{i_2}^{(k')})\\
\frac{1}{2}(f_{i_1}^{(k')}-f_{i_2}^{(k')})
\end{array}\right)
=\left(\begin{array}{c}
{A_i}^{(k)}\\
{B_i}^{(k)}\\
{f_i}^{(k)}\\
{r_A}^{(k)}\\
{r_B}^{(k)}\\
{r_f}^{(k)}
\end{array}\right).
\end{eqnarray*}
Note that the determinant of the Jacobian of the transformation
$\textrm{t}_{k\mapsto k'}$, (i.e.\ the determinant of the above
6-by-6 matrix) is $|J_{k\mapsto k'}|=2$, and that of its inverse is
$1/2$. Thus, the acceptance probability for increasing a model by
one signal is given by
\begin{equation}
\alpha_{k  \mapsto k'}(\avec_{k'}'|\avec_{k}) =\min \left\{ 1,
\frac{p(k', \avec_{(i_1)}',\avec_{(i_2)}')
p(\dvec|\avec_{(i_1)}',\avec_{(i_2)}',k') }{p(k,\avec_{(i)})
p(\dvec|\avec_{(i)},k) q(\rvec) } \right\} \left| J_{k  \mapsto
k'} \right|.
\end{equation}
By analogy, the acceptance probability for the reversible move of
a fusion of two signals is  given by
\begin{equation}
\alpha_{k' \mapsto k}(\avec_{k}|\avec_{k'}') =\min \left\{
1,\frac{p(k,\avec_{(i)}) p(\dvec|\avec_{(i)},k) q(\rvec) }{p(k',
\avec_{(i_1)}',\avec_{(i_2)}')
p(\dvec|\avec_{(i_1)}',\avec_{(i_2)}',k') } \right\}  \left| J_{k'
\mapsto  k} \right|,
\end{equation}
where $|J_{k'\mapsto
k}|=\left|\frac{\partial(\avec_{(i)},\rvec)}{\partial(\avec_{(1i)}
', \avec_{(2i)}')}\right|$.

We use a multivariate normal proposal distribution,
$N[\textbf{0},\textrm{diag}(\sigma_A^2,\sigma_B^2,\sigma_f^2)]$,
for $q(\rvec)$. Suitable values for the variances of this
multivariate normal distribution can be chosen by considering
their effect on the acceptance probabilities. First, we consider
the effect of a small proposal variance for a splitting
transition. In this case the proposal has an insignificant effect
on the likelihood since the two new signals in the model function
are almost linearly dependent. On the other hand, the resulting
large value for $q(\rvec)$  considerably decreases the acceptance
probability when proposing a split, and increases it when a fusion
of two signals is proposed. Now we consider the effect of a large
proposal variance: the value of $q(\rvec)$, and therefore its
influence on the acceptance probability, is moderate. However it
causes the likelihood to change considerably, resulting in a small
acceptance probability. The choice of $\sigma_A^2, \sigma_B^2$ and
$\sigma_f^2$ is therefore an important consideration for improving
mixing. In each iteration we set $\sigma_A^2, \sigma_B^2$ equal to
the noise level $\sigma^2_m$ of the current model $m$.

The posterior precision of the frequency in a single-frequency
model depends on the signal to noise ratio
$\gamma=\sqrt{(A^2+B^2)/\sigma^2}$ and the number of samples $N$
of the data set \cite{Jaynes87}. Using a Gaussian approximation to
the posterior pdf of the frequency \cite{Jaynes87}, its standard
deviation is given by $\sigma_{f}''={(2\pi\gamma)}^{-1}\sqrt{48 /
N^3}$ . This yields a distance in frequency for which two
sinusoids can still be identified as distinct, neglecting the
interference of other sinusoids. We therefore use  $\sigma_{f}''$
as a frequency perturbation when splitting two sinusoids and use a
normal distribution with this particular standard deviation when
merging.

\subsection*{Birth-and-death transitions}
A `birth' transformational step simply creates a new signal with
parameter triple $\avec_{(i)}'$ independent of other existing
signals in the current model $\mathcal{M}_{k}$. The one-to-one
transformation in this case is very simply given by
$\textrm{t}_{k\mapsto k'}(\rvec)=\rvec =\avec_{(i)}'$.   The inverse
(`death') transformation  that annihilates signal $i'$,
$\textrm{t}^{-1}_{k\mapsto k'}:=\textrm{t}_{k'\mapsto k}$, has form
$\textrm{t}_{k'\mapsto k}\left( \avec_{(i)}'
\right)=\avec_{(i)}'=\rvec$.  The Jacobian for both of these is 1.
The acceptance probability for the creation process is therefore
\begin{equation}
\alpha_{k \mapsto k'}(\avec'_{k'}|\avec_k)=\min \left\{ 1, \frac{
p(k') p(\avec_{(i)}') p(\dvec|\avec'_{k'},k') }{p(k)
p(\dvec|\avec_k,k) q(\rvec) } \right\} ,
\end{equation}
and that for the annihilation process is
\begin{equation}
\alpha_{k' \mapsto k}(\avec_k|\avec'_{k'}) = \min \left\{ 1,
\frac{p(k) p(\dvec|\avec_k,k) q(\avec_{(i)}') }{ p(k')
p(\avec_{(i)}') p(\dvec|\avec'_{k'},k') } \right\}.
\end{equation}

As for split and merge transitions, a bold proposal distribution $q$
results in a small acceptance probability due to the strong effect
on the likelihood, whereas timid proposals have minor effects on the
likelihood but are often rejected due the higher values of the
proposal distribution. An effective way to make a proposal for the
frequencies is to base it on the Schuster periodogram of the data
\cite{schuster05}, given by
\begin{equation}
C(f)=\frac{1}{N} \left[ R(f)^2+I(f)^2 \right], \label{periodogram}
\end{equation}
where $R(f)=\sum_{j=1}^N{d_j \cos \left( 2 \pi f t_j\right)}$ and
$I(f)=\sum_{j=1}^N{d_j \sin \left( 2 \pi f t_j\right)}$ are the real
and imaginary parts from the sums of the discrete Fourier
transformation of the data.
This technique has already been applied successfully
by \cite{Andrieu99}. We use proposals from a normal distribution for
the amplitudes with zero mean and a standard deviation derived from
the average amplitude of all remaining amplitudes of the current
state of the Markov chain.

Classical MCMC methods could be used for transitions within a
particular model $\mathcal{M}_m$, however we use an adaptive MCMC
technique here. The delayed rejection (DR) method has been
introduced by \cite{TierneyMira99, Mira98,GreenMira01} and several
of us have successfully applied it to estimate the frequency and
frequency derivative of potential gravitational radiation signals
produced by a triaxial neutron star \cite{pul2}.

\subsection{The delayed rejection method for parameter estimation}
\label{sectionDL} As sampling progresses, suppose that at the
$n$th iteration the state of the Markov chain is $\avec=\avec_m$
from model $\mathcal{M}_{m}$. We can choose a new state within the
same model by first sampling a candidate state $\avec'$ from a
proposal distribution $q_1(\avec'|\avec)$ and then accepting or
rejecting it with an MH-probability $\alpha_1(\avec'|\avec)$
depending on the distribution of interest. When a proposed MH move
is rejected, a second candidate $\avec''$ can be sampled with a
different proposal distribution $q_2(\avec''|\avec',\avec)$ that
may depend on the previously rejected proposal. To preserve
reversibility of the Markov chain and thus to comply with the
detailed balance condition, the acceptance probabilities for both
the first and the second stage are given by \cite{Mira98}
\begin{equation}
\alpha_1(\avec '|\avec) =
 \min\left[1,\frac{p(\avec')p(\dvec|\avec')
 q_1(\avec|\avec')}{p(\avec)p(\dvec|\avec)q_1(\avec'|\avec)}\right]
\end{equation}
and
\begin{equation}
\alpha_2 (\avec''|\avec',\avec) = \min\left\{1,\frac
{p(\avec'')p(\dvec|\avec'') q_1(\avec'|\avec'')
q_2(\avec|\avec',\avec'')[1
-\alpha_1(\avec'|\avec'')]}{p(\avec)p(\dvec|\avec)
q_1(\avec'|\avec)
q_2(\avec''|\avec,\avec')[1-\alpha_1(\avec'|\avec)]}\right\}.
\end{equation}
We therefore apply distinct types of DR transition for the
amplitudes and the frequency of a sinusoid, and these are considered
below. The transitions are performed randomly and with equal
probability for a randomly chosen sinusoid $i$.

\subsection*{Proposing new amplitudes}
Our amplitude proposal distribution is multivariate Normal, with a
covariance matrix that depends on the basis functions of the
sinusoids. The closer two sinusoids are in frequency the more
correlation there is in their recovered amplitudes.

In \cite{Andrieu99} the amplitudes were regarded as nuisance
parameters and integrated out by treating them as parameters of a
multiple regression model with a conditional posterior that is
normally distributed with known mean and covariance matrix according
to the $g$-prior. If deemed necessary the amplitudes can be updated
by Gibbs sampling. However, the computation of the covariance matrix
involves determining the inverse of
$\D^{(m)}(\fvec)^{T}\D^{(m)}(\fvec)$ for each iteration, which is
impractical considering the number of sinusoids expected for LISA.
Mindful of this we consider the interference between the sinusoids
by proposing those pairs of sinusoids that have the strongest
correlation, namely sinusoid pairs which are closest in frequency.
Therefore for the randomly chosen sinusoid $i$ we also involve the
sinusoid $i_{c}$ which is the closest in frequency. The matrix for
this subset of basis vectors reduces to
\begin{equation}
\D_{(i,i_{c})}^{(m)}(\fvec)=\left(\begin{array}{c c c c} \cos(2\pi
f^{(m)}_{i} t_1) & \sin(2\pi f^{(m)}_{i}  t_1) & \cos(2\pi
f^{(m)}_{i_{c}}  t_1) &
\sin(2\pi f^{(m)}_{i_{c}} t_1) \\
\cos(2\pi f^{(m)}_{i} t_2) & \sin(2\pi f^{(m)}_{i}  t_2) & \cos(2\pi f^{(m)}_{i_{c}}  t_2) &
\sin(2\pi f^{(m)}_{i_{c}} t_2) \\
\vdots & \vdots & \vdots & \vdots \\
\cos(2\pi f^{(m)}_{i}  t_N) & \sin(2\pi f^{(m)}_{i} t_N) &
\cos(2\pi f^{(m)}_{i_{c}}  t_N) & \sin(2\pi f^{(m)}_{i_{c}} t_N),
\end{array} \right)
\end{equation}
which results in a covariance matrix merely of type $4 \times 4$ for
computing proposals for a sinusoid pair $(i,i_{c})$. Now suppose
that
$\avec_{(i,i_{c})}=\left(A^{(m)}_{i},B^{(m)}_{i},A^{(m)}_{i_{c}},B^{(m)}_{i_{c}}
\right)^{T}$ is a vector containing the amplitudes of the sinusoid
pair (but not their frequencies). Then
\begin{equation}
\Sigma_{(i,i_{c})}=r \left[\avec^{T}_{(i,i_{c})} \left(
\D_{(i,i_{c})}^{(m)}(\fvec) \right)^{T}
\left(\D_{(i,i_{c})}^{(m)}(\fvec) \right) \avec_{(i,i_{c})} \right]
\left( \D_{(i,i_{c})}^{(m)}(\fvec)^{T}\D_{(i,i_{c})}^{(m)}(\fvec)
\right)^{-1}
\end{equation}
is the covariance matrix of the proposal distribution
$q_1(\avec'_{(i,i_{c})}|\avec_{(i,i_{c})})=N(\avec_{(i,i_{c})},\Sigma_{(i,i_{c})})$,
where $r$ is a factor controlling the step size of the proposal.

If the first proposal is rejected, another attempt is made by
proposing the amplitudes of a single sinusoid $i$ without
considering correlations. The variance of the proposal is assessed
by the mean amplitude of all signals which is given by
$\bar{a}^{(m)}(\avec)=(2m)^{-1}\sum_{i=1}^m{[(A^{(m)}_i)^2+(B^{(m)}_i)^2]}$.
This yields the second proposal
$q_2(\avec''_{(i)}|\avec)=N\left[\avec_{(i)},r \cdot \textrm{diag}
\left(\bar{a}^{(m)}(\avec),\bar{a}^{(m)}(\avec)\right)\right]$
where $\avec_{(i)}=\left(A^{(m)}_{i},B^{(m)}_{i}\right)^{T}$ is
the subvector of $\avec$ containing the amplitudes of sinusoid
$i$.

\subsection*{Proposing a new frequency}
A new frequency is proposed as follows: In the first stage a new
frequency for sinusoid $i$ is sampled from a proposal density that
is proportional to the periodogram $q_1(\avec'_{(i)})\propto
\left(0,0,C(f) \right)^T$ and independent of the actual stage.
This is similar to the sinusoid proposal scheme for RJMCMC. The
main objective of this stage is to coarsely scan the whole
parameter space for frequencies.

The event of a rejection suggests to sample from the local
frequency mode and a proposal is made conditional on the actual
state of the frequency by slightly perturbing the state. In the
same manner as in the split-and-merge transition the perturbation
is oriented on the possible achievable accuracy
$\sigma_{f^{(m)}_{i}}={(2\pi)}^{-1}\sqrt{48 \sigma_{m}^2  \left[
(A^{(m)}_{i})^2 + (B^{(m)}_{i})^2\right]^{-1} N^{-3}}$ of a
frequency by \cite{Bretthorst88}. This yields the proposal
$q_2(\avec''_{(i)}|\avec_{(i)})=N\left(\avec_{(i)},\textrm{diag}\left(0,0,
\sigma_{f^{(m)}_{i}}\right)\right)$ and aims to draw
representative samples from the local mode in the second stage.

\subsection{Updating the noise parameter}
\label{updating_the_noise} The sum of the squared residuals
between the model and the data, taken from the likelihood in
Eq.~\ref{LH}, is
\begin{equation}
S^2=\sum_{j=1}^N{\left[d_j - s_m(t_j,\avec_m)\right]^2}.
\end{equation}
Using this, we choose a vague prior for the noise parameter
$\sigma^2_m$, defined by $IG(\alpha,\beta)=IG(N_p/2,N_p \cdot
S_p^2/2)$ with a shape parameter $\alpha=N_p/2=0.001$ and a scale
parameter $\beta=N_p S_p^2/2=0.001$.  This yields $N_p=0.002$ and
$S_p^2=1$ for the parameters of the vague prior. The full
conditional distribution
\begin{equation}
p(\sigma_m^2|m,\avec_m,\dvec)\propto
IG\left(\frac{N_p+N}{2},\frac{N_p S^2_p + S^2} {2}\right)
\end{equation}
is used for drawing samples for $\sigma^2_m$ in a Gibbs update.

\subsection{Initial values}
The initial values of a Markov chain are crucial for the length of
the burn-in period needed to converge to the real posterior
distribution. We could start with and empty model, $\mathcal M_0$,
but it is obvious that it would then take the sampler many steps
to find all the signals. Instead, we perform a Fast Fourier
Transformation (FFT) of the data and use this to generate our
initial values. We use the frequencies corresponding to the local
maxima in the periodogram, $f_{0,i}$, as starting values for
$f_{0,i}^{m_0}$ and $A_{0,i}=2 R(f_{\textrm{max}_i})/N$,
$B_{0,i}=2 I(f_{\textrm{max}_i})/N$ as starting values for
$A_{i}^{m_0}$ and $B_{i}^{m_0}$, respectively. Theoretically we
could use all the local maxima as initial values, but as most of
them are due to noise we select only those that exceed a certain
threshold. We set this threshold low, as it is easier to delete
non-relevant sinusoids than create good ones.

The frequency resolution depends on sample size $N$, suggesting that
the convergence of the Markov chain is also dependent on $N$. As we
will see, spectral estimates based on the FFT are significantly
worse than those we obtain by our MCMC method, but they are
sufficient to serve as initial values.

\subsection{Identifying the sinusoids}
\label{classify} Although the RJMCMC method enables us to select
the most probable model, we still encounter the
\emph{label-switching problem}.  This is a general problem caused
by the invariance of the likelihood under relabelling of the
sinusoidal components, and has been extensively discussed in the
context of mixture models \cite{Celeux00}. During the MCMC
simulation, parameter triples are constantly changing their
affiliation to individual sinusoids, either due to the creation
and annihilation of sinusoids or following transitions within the
same model. We therefore need an additional step in our analysis
if we are to break this symmetry and talk meaningfully about
individual sinusoid components.  This step involves associating
the samples in the final Markov chain with particular sinusoids,
which we know neither by number nor by location.

The sinusoids that are contained in the model $\hat{\mathcal M}$,
with the highest posterior probability of $m$, are permutations of
$\hat{m}$ coexistent sinusoids from a list of unknown size.  We
will assume therefore that this list is the same size as the upper
limit $m_{\ rm max}$ of the marginal posterior of $m$. However, in
practice the coexistent sinusoids in $\hat{\mathcal M}$ are
usually clear, and it is rather unlikely that others present in
higher order models could replace them. The parameter vector
sampled in each iteration of the Markov chain (corresponding to
model $\hat{\mathcal M}$) is a permutation of $\hat{m}$ parameter
triples determining $\hat{m}$ out of $m_{\rm max}$ sinusoids. The
problem is to determine which parameter triple belongs to which
sinusoid. We find that the parameter that contributes
significantly to identifying a sinusoid is its frequency. Thus in
order to obtain the dominant sinusoids that characterize model
$\hat{\mathcal M}$ we calculate the marginal posterior of the
frequency and obtain the $\hat{m}$ strongest peaks together with
their frequency ranges by finding the threshold that separates
those peaks.

Since we have to deal with a vast number of output samples grouped
within very small regions we cannot apply kernel density estimates
or histograms, as the required fixed bin size for a histogram
would be too small to be feasible. Instead, we use a variable bin
size and calculate densities for a fixed number of samples per
bin. We initially sort all individual parameter triples for all
MCMC samples of the considered model $\hat{\mathcal M}$ by
frequency. If we have generated $n$ MCMC samples of model
$\hat{\mathcal M}$ during a run then we have to deal with $\hat{m}
n$ parameter triples and hence $f_1<\ldots<f_{n\hat{m}}$ ordered
frequency samples. The density can be assessed by calculating the
frequency range spanned by a fixed number of sorted frequencies.
The $\hat{m} n$ parameter triples are from $\hat{m}$ sinusoids and
therefore we can expect $n$ frequency samples per peak since we
assume the number of peaks to be similar to the number of
sinusoids. Hence the fixed number of sorted frequencies that spans
the frequency ranges must be some fraction, $r$, of the number of
MCMC samples $n$, and is the counterpart to the required bin width
in a histogram or the bandwidth $h$ of a kernel density estimate
$\textrm{f}(f)=(2hn\hat{m})^{-1}\sum_{i=1}^{n
\hat{m}}{\textrm{I}_{|f-f_i|<h}}$ with uniform kernel, where
$\textrm{I}$ is the indicator function. The advantage of this
approach is the automatic adaptation of the bandwidth to the
situation by involving the information of the expected parameter
triples per peak. We found $r=0.05$ (5\%) to be a good value, and
hence $r n$ serves as an estimate of the number of samples needed
for assessing the spans of the frequency ranges. In analogy to the
kernel density estimate with uniform kernel, where the number of
samples are counted that fall into a range of length $2h$, the
density value $\rho_j$ corresponds to each frequency sample $j$
and its $rn-1$ subsequent samples that fall into a frequency range
of length $\left(f_{j+rn-1}-f_{j}\right)$. Hence the density is
given by $\rho_j =rn/\left[n\hat{m} \left(f_{j+rn-1}-f_{j}\right)
\right]= r/\left[\hat{m}(f_{j+rn-1}-f_{j}) \right]$ where
$j={1,\cdots,n(\hat{m}-r)}$. Since each $\rho_j$ comprises the
samples ${j,\cdots,j+rn-1}$ this has to be considered later when
deriving spans for the peaks. We find the smallest density
threshold $l$ that separates $\hat{m}$ distinct peaks with respect
to the values of $\rho_j$. The frequency range for each peak $k
\in \{1,\cdots,\hat{m}\}$ is $[f_{j_{k,{\rm start}}},f_{j_{k,{\rm
end}}+rn-1}]$ where $j_{k,{\rm start}}$ and $j_{k,{\rm end}}$ are
the indices of the first and the last member of the set of
$\rho_j$'s in peak $k$, respectively. Due to the fact that we
always focus on frequency ranges that contain a fixed amount of
frequency samples we efficiently deal with large frequency ranges
of low density. This technique is fast and requires a minimum of
memory. The greatest computational cost is in sorting the
frequencies, although this can be carried out fairly quickly using
a heap sort.

It is still possible that individual peaks contain more than one
sinusoid or even none. This can be assessed by the histogram of
the number of sinusoids in the MCMC samples for the restricted
frequency range under consideration. To separate more than one
present sinusoid, we then consider the two amplitudes and apply an
agglomerative hierarchical cluster analysis that involves all
three parameters. We use a modified Ward technique \cite{Ward63}
that minimizes the within-cluster variance using a normalized
Euclidean distance between the parameters by adjusting the
frequency range to the much larger range of the amplitudes. We use
the software package R \cite{R} for this task. The Ward technique
starts with each parameter triple belonging to a singleton
cluster. Iteratively, cluster pairs are joined that produce the
smallest possible increase in within-cluster sum of squares. In
this particular case we have no difficulty in detecting when to
stop the agglomeration, which is an important issue in cluster
analysis, as we know the expected number of sinusoids in a peak
and hence the number of clusters. However, due to the vast number
of parameter triples involved here it is not possible to carry out
a cluster analysis simultaneously on all of the data. Instead we
have to divide the set of samples into randomly chosen subsets of
equal size and perform separate cluster analyses (with R) for each
of these subsets. Finally, we perform a cluster analysis on the
median points of the subset clusters to allocate each of those
clusters to a super-cluster. Each single parameter triple is then
allocated to a super-cluster and hence to a presumed sinusoid.
Those parameters which fall into small peaks below the threshold
can be allocated to an additional noise contribution.

%% ************************ RESULTS **************************

\section{Simulations}
\label{simulations} 
\begin{figure}[hbt]
\centerline{\includegraphics[width=14cm]{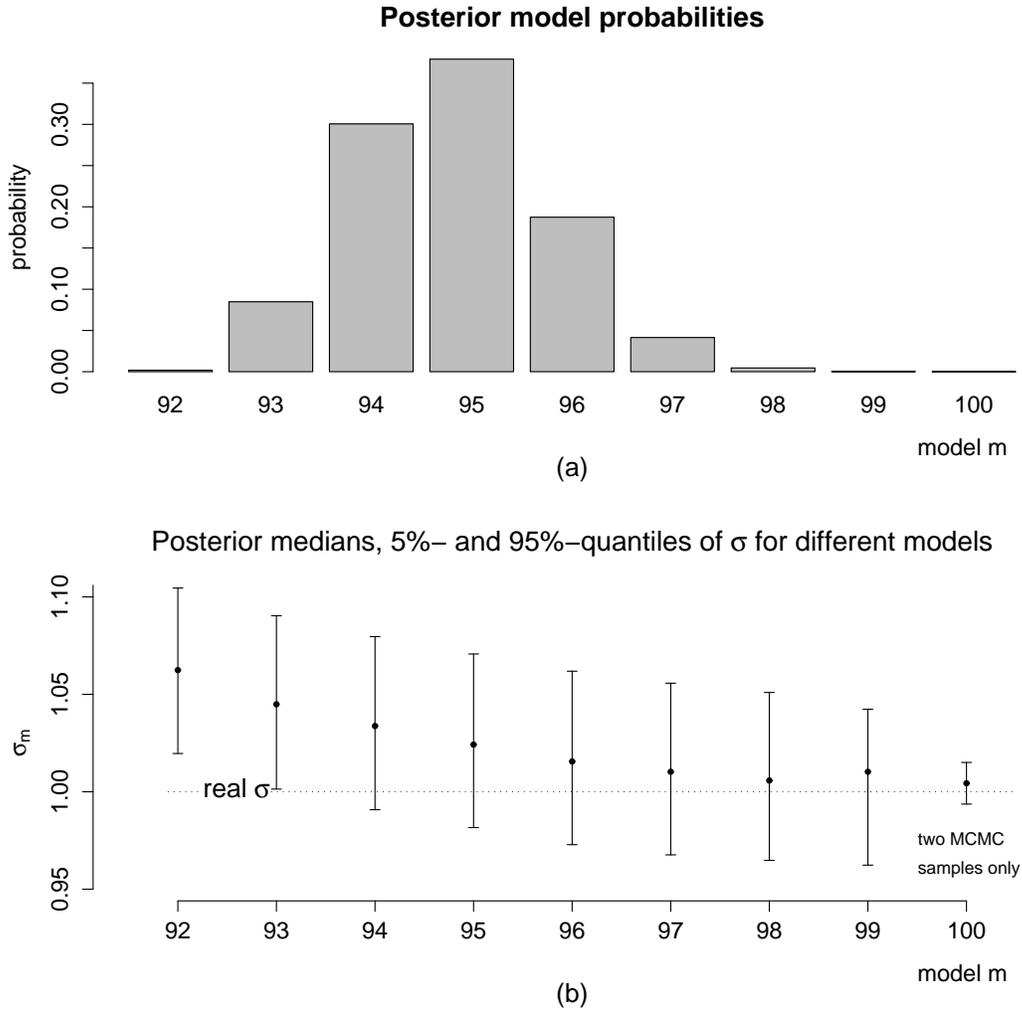} } \caption{(a)
shows the model probabilities deduced from our analysis. The model
corresponding to 95 sinusoids has highest probability. Each model
has a different noise level, and (b) shows the estimated noise
standard deviation and their confidence intervals for the different
models. The dotted line indicates the real noise level. As model 100
has a very low probability only two MCMC were obtained during the
run of $3 \times 10^8$ iterations, making it poorly determined.}
\label{fig_2}
\end{figure}

We created an artificial data set of 1\,000
samples containing  $100$ random sinusoids in gaussian noise. The
amplitude coefficients were chosen randomly in the range
$[-1\ldots 1]$ and the noise standard deviation was $\sigma=1$,
making our maximum signal to noise ratio,
$\gamma_i=\sqrt{(A_i^2+B_i^2)/\sigma^2}$, about $1.4$. We will
present our results with dimensionless units. We chose a uniform
prior for $m$ on $\{0,1,2, \ldots, M=60\,000\}$, and set $A_{\rm
max}=B_{\rm max}=5$. The Markov chain ran for $3\times 10^8$
iterations, the first $5 \times 10^6$ of which were considered as
burn-in and discarded. The chain was then thinned by storing every
$2\,000$th iteration. The MCMC simulation was implemented in C on
a 2.8\,GHz Intel P4 PC and took about 78 hours to run.
Fig.~\ref{fig_2}(a) gives the histogram of the marginal posterior
model probabilities obtained by the reversible jump algorithm. As
each model $\mathcal{M}_m$ is characterized by a different noise
level $\sigma_m$, we have also plotted the marginal posterior
distributions of the noise standard deviations in
Fig.~\ref{fig_2}(b).
Note that $\sigma_m$ decreases with higher model order $m$ since a
model comprising more sinusoids accounts for more of the available
power.
%However, the highest posterior model probability $m=95$ does not
%necessarily imply the presence of 95 signals but the presence of 95
%coexistent signals permuted out of a set of $m_{tot}$ signals, where
%$m_{tot}$ is unknown. Hence the largest observed model order $m=100$
%of the posterior pdf of $m$ as shown in Fig.~\ref{fig_2} indicates
%the total number of signals that may be present. Furthermore, the
%range of the histogram, here from 92-100, indicates how many of
%these signals are too weak for detection or too close to be
%separated.

All subsequent results presented here are based solely on MCMC
samples corresponding to model ${\cal M}_{95}$, so we will omit
the superscripts and denote the parameter vector of model ${\cal
M}_{95}$ by
$(A_1,B_1,f_1,\ldots,A_{95},B_{95},f_{95},\sigma_{95}^2)$. The
power spectral density of the signal can be estimated from the
product of the conditional expectation of the energy of each
sinusoid $i$ given its frequency $f_i$,
$E(A_i^2+B_i^2|\dvec,m,f_i)$, and the posterior pdf of $f_i$ given
the data, $p(f_i|\dvec)$ as described in \cite{Jaynes87}. An
advantage of using a Bayesian spectral analysis is the subsequent
ability to calculate confidence areas for the spectrum by grouping
our MCMC samples and calculating posterior confidence intervals
for frequency bins. An appropriate width for the bins can be
estimated using the frequency resolution of the spectrum,
$\sigma_{f}=(2\pi\gamma)^{-1} (48 / N^3)^{1/2}$, given by
\cite{Bretthorst88}. In this example the choice of 20\,000 bins is
sufficient to resolve sinusoids of $\gamma\le 1.4$.

\begin{figure}[hbt]
\centerline{\includegraphics[width=16cm]{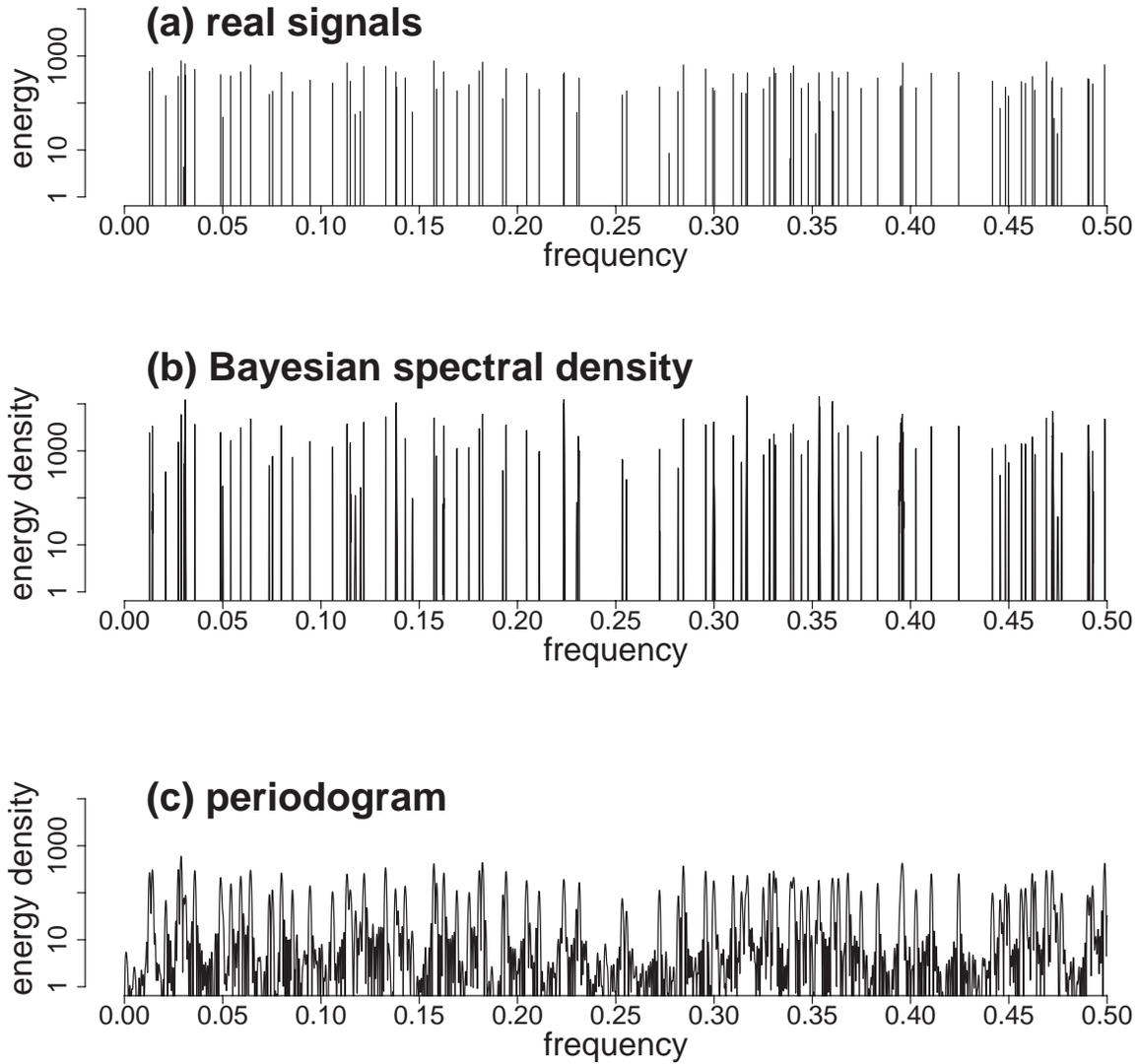} }
\caption{Comparison of (a) the real signal energies, (b) the
Bayesian spectral density, and (c) the classical periodogram of
the test data. Note that (a) shows component energies, whereas (b)
and (c) show energy densities (inferred energy per unit
frequency). The peaks generated by the Bayesian method are higher
than those of the periodogram, reflecting its greater ability to
constrain the frequency. } \label{fig_3}
\end{figure}

Fig.~\ref{fig_3} compares the known real sinusoids from which the
data set had been created, the Bayesian spectral density estimate
as described above, and the classical periodogram derived from
Eq.~\ref{periodogram}. We note here that Fig.~\ref{fig_3}(b) and
Fig.~\ref{fig_3}(c) are spectral densities, whereas
Fig.~\ref{fig_3}(a) is a plot of the component energies, defined
as $N(A_i^2+B_i^2)/2$.  The theoretical spectral density of these
components would consist of delta functions of formally zero width
and hence infinite density. Furthermore the accuracy of the
Bayesian spectral estimate is better than the one obtained from
the periodogram, which also also leads to different values on the
ordinate. The Bayesian spectrum estimate usually reveals much
higher values in the density since the same energy contribution
from a sinusoid is distributed over a smaller frequency range.
However the locations of the peaks can be compared directly, which
is the main purpose of Fig.~\ref{fig_3}.

\begin{figure}[hbt]
\centering
\begin{tabular}{c|c|c}
\includegraphics[width=5.5cm]{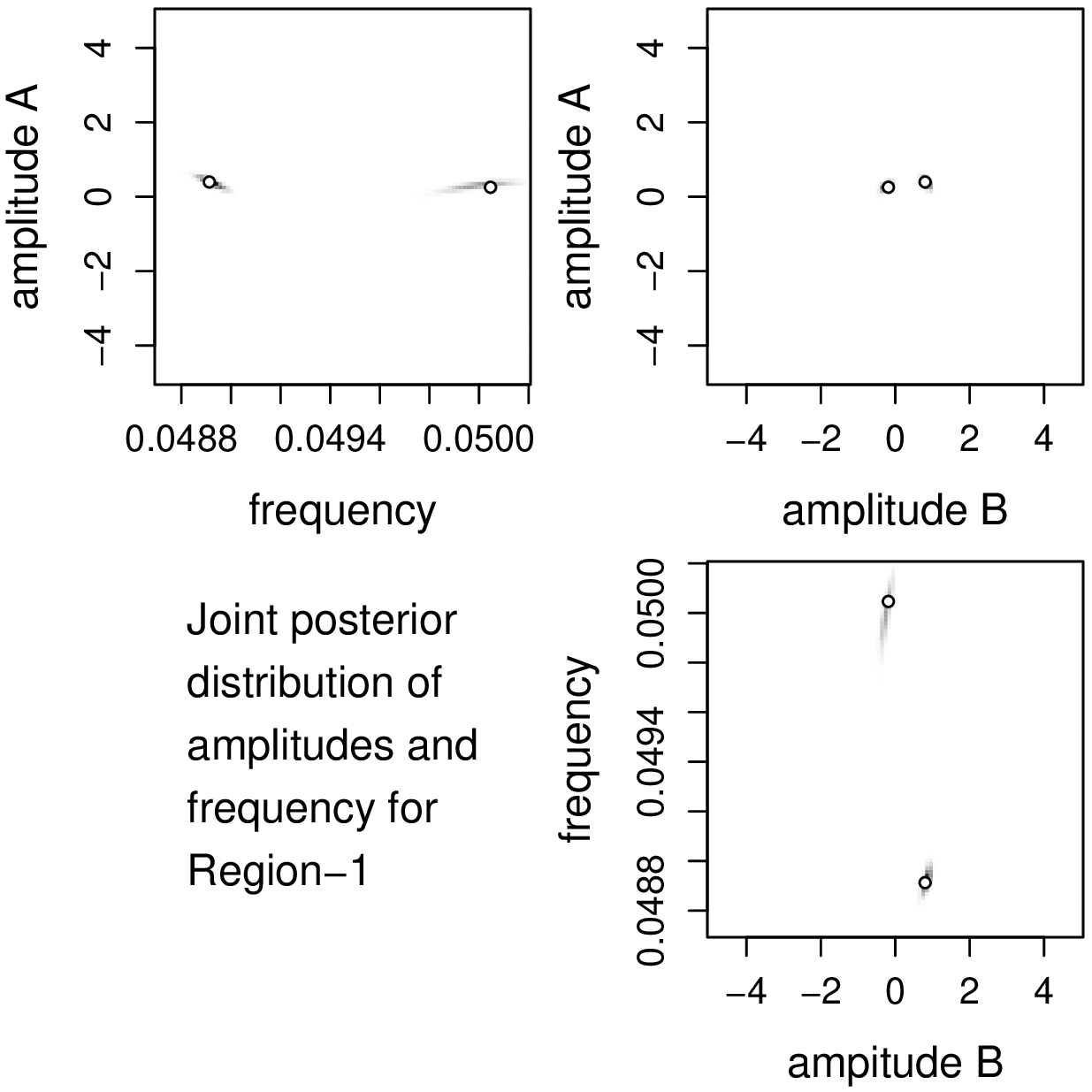}  &
\includegraphics[width=5.5cm]{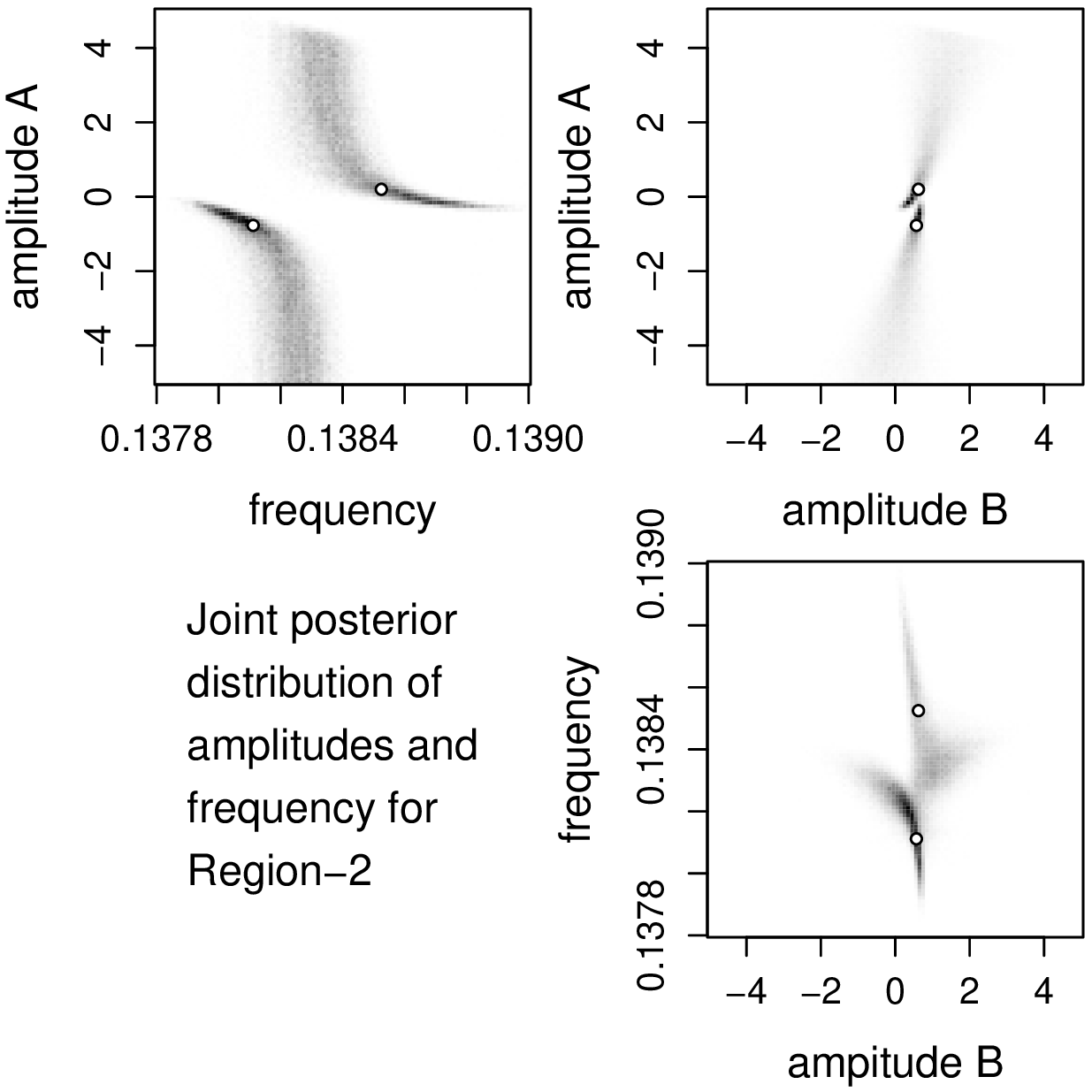}  &
\includegraphics[width=5.5cm]{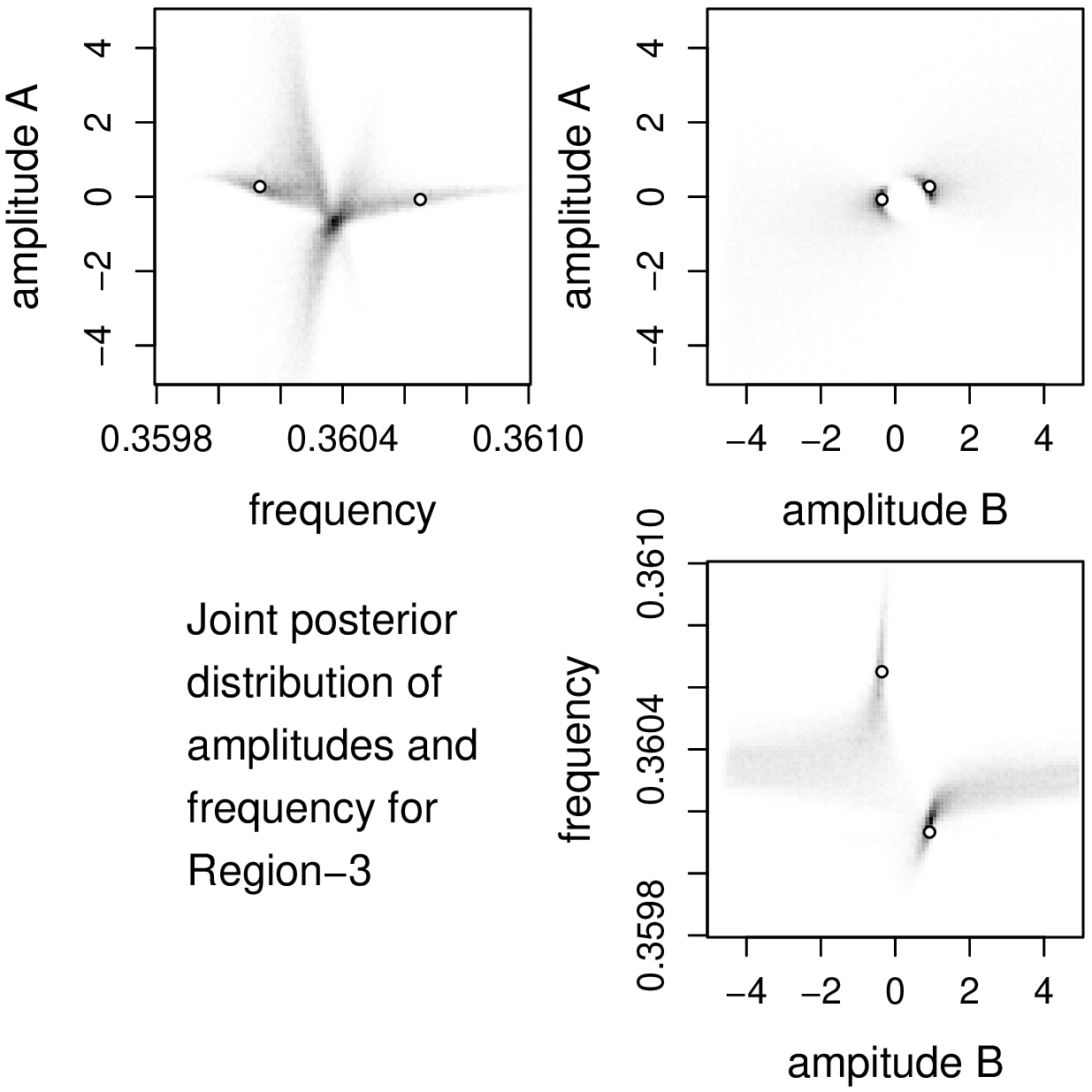}
\end{tabular}
\caption{Joint posteriors for the amplitudes and frequency in
Regions-1, -2 and -3. The circles show the \emph{true} values of the
parameters comprising signal pairs.} \label{fig_4}
\end{figure}

We have picked out three different frequency ranges for special
attention. Each contain two close sinusoid pairs and we will refer
to these as Region-1:$f \in [0.0486,0.0506]$, Region-2:$f \in
[0.1375,0.1395]$, Region-3:$f \in [0.3594,0.3614]$. Joint posteriors
for these regions are presented in Fig.~\ref{fig_4}.

In Region-1 we see two sinusoids with a frequency gap of
approximately $1/N=0.001$ which, for a discrete Fourier transform,
is the size of a single frequency bin. Region-2 and Region-3 contain
sinusoids with smaller frequency gaps of $0.000\,413$ and
$0.000\,517$, respectively. The posteriors for Region-1 are strongly
peaked at the true values. The same is true for Region-2 and
Region-3, but we observe a larger uncertainty in their frequencies
and even more in their amplitudes especially when the frequencies of
the sinusoids are very close.  The reason for this is clear: when
the frequencies of the sinusoids are very close together  they
slowly beat against each other, and for these particular signals the
beat period is greater than the observing time so that their total
energy is difficult to assess.

\begin{figure}[hbt]
\centerline{ \includegraphics[width=16cm]{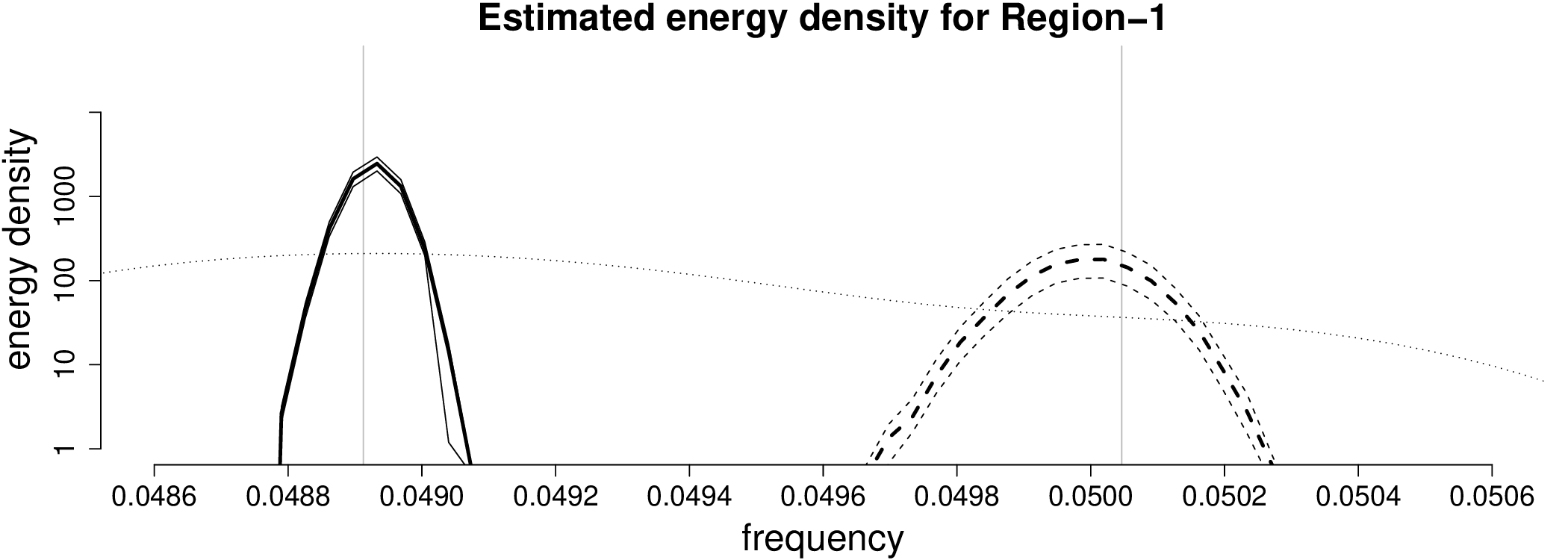} }
\vspace{0.5cm} \centerline{
\includegraphics[width=16cm]{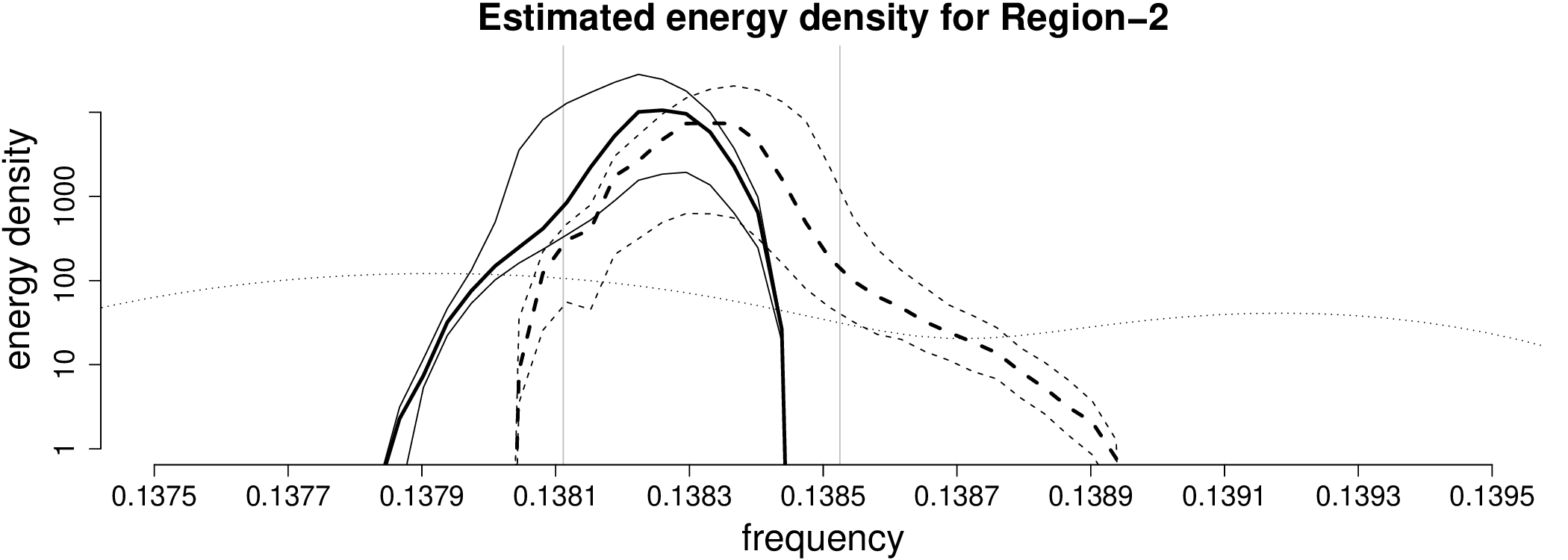} } \vspace{0.5cm}
\centerline{ \includegraphics[width=16cm]{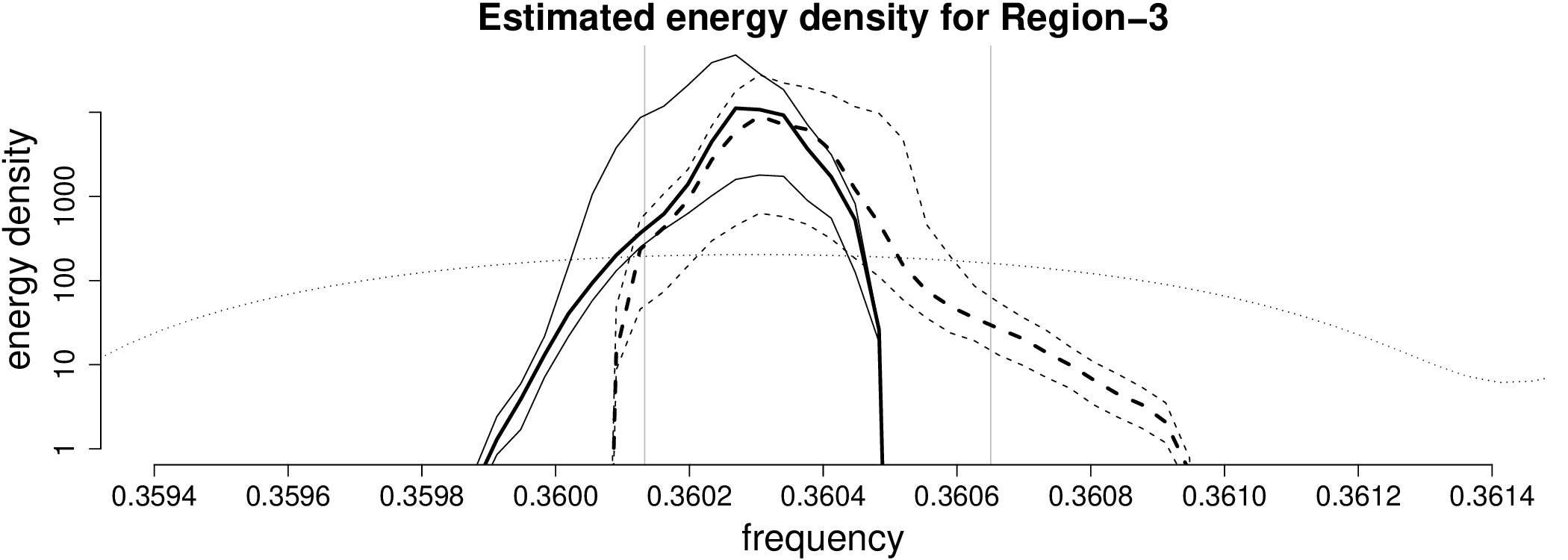} }
\caption{Bayesian spectral estimates, showing 95\% confidence
bounds in spectral density, for the three regions considered. Each
regions contain two distinct sinusoids, one shown with solid lines
and the other with dashed lines. The three lines for each sinusoid
show the median value (thick) and 95\% bounds (thin) of the
spectral estimate at each frequency. The vertical lines show the
true frequency values of these components.  The dotted line
traversing the entire width of the plots is the corresponding
classical periodogram. } \label{fig_5}
\end{figure}

The spectral estimates for these regions are shown in greater detail
in Fig.~\ref{fig_5}.  Rather than scale the frequency posterior by
the expectation of the energy, as in Fig.~\ref{fig_3}, here we scale
by the 2.5\%, 50\% and 97.5\% quantiles of the energy for each
frequency bin to indicate the uncertainty in the value of the
spectral density.

The gap between the two sinusoids in Region-1 of Fig.~\ref{fig_5}
is about $1/N$, and the sinusoid with the lower frequency has a
significantly larger energy than its neighbor, so its frequency is
inferred with greater accuracy.   Region-2 and Region-3 however
comprise pairs of considerably closer sinusoids, making it
considerably more difficult to infer their separate frequencies.
The corresponding uncertainties in their amplitudes can be seen in
both Fig.~\ref{fig_4} and Fig.~\ref{fig_5}, which shows large
energy density values, especially in the area where the
frequencies overlap.

\begin{figure}[hbt]
\centering
Region-1
\includegraphics[width=15cm]{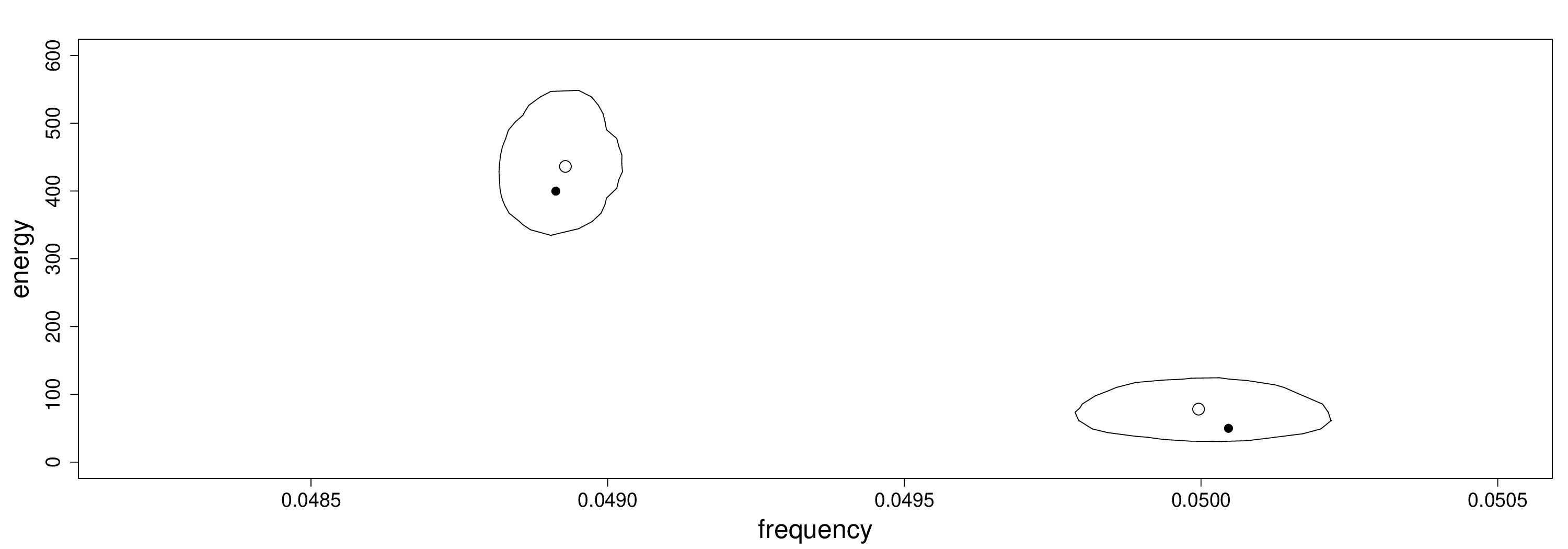}
\begin{tabular}{p{8cm}p{8cm}}
\center Region-2 & \center Region-3
\end{tabular}
\begin{tabular}{p{8cm}p{8cm}}
\includegraphics[width=7.5cm]{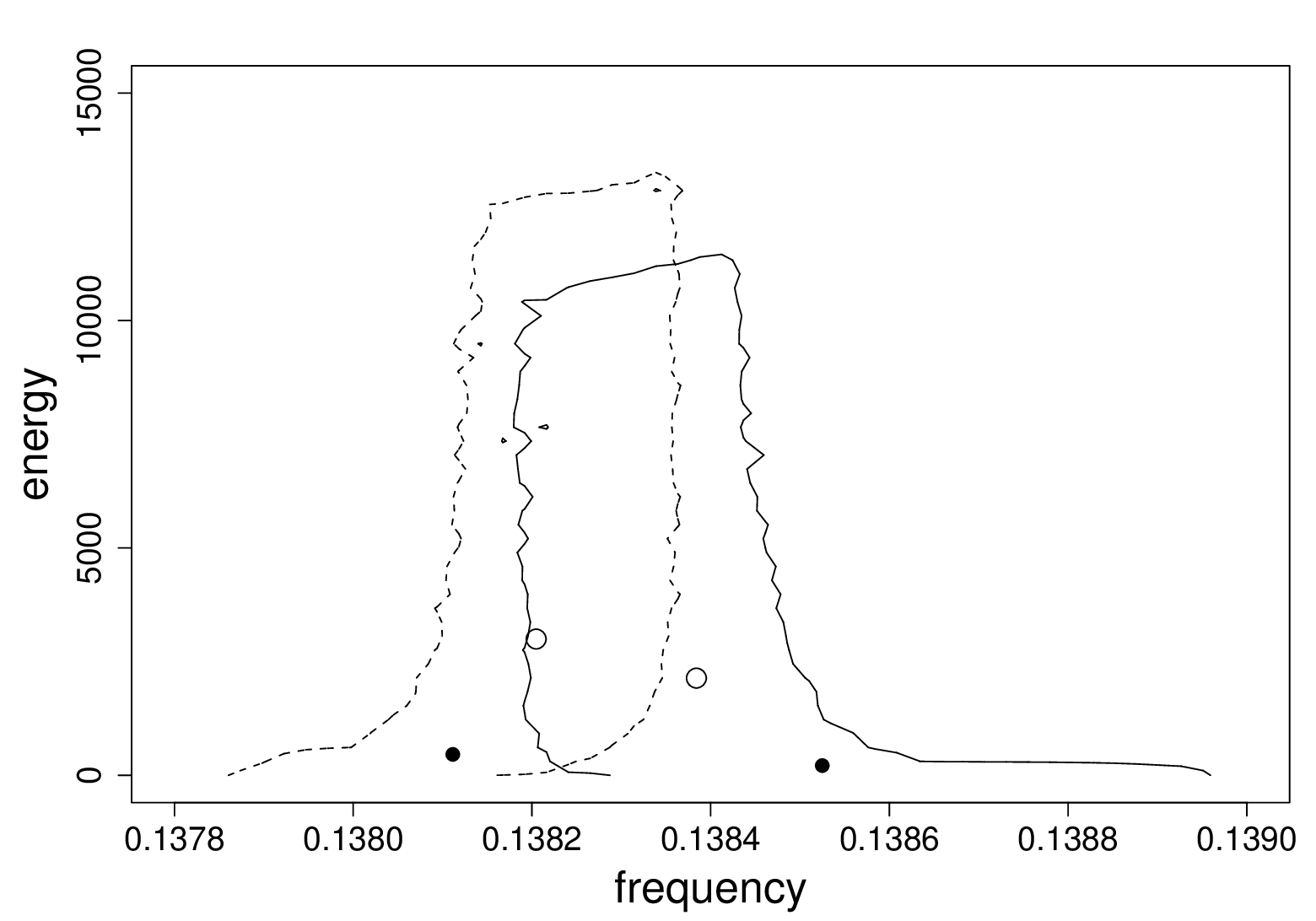} &
\includegraphics[width=7.5cm]{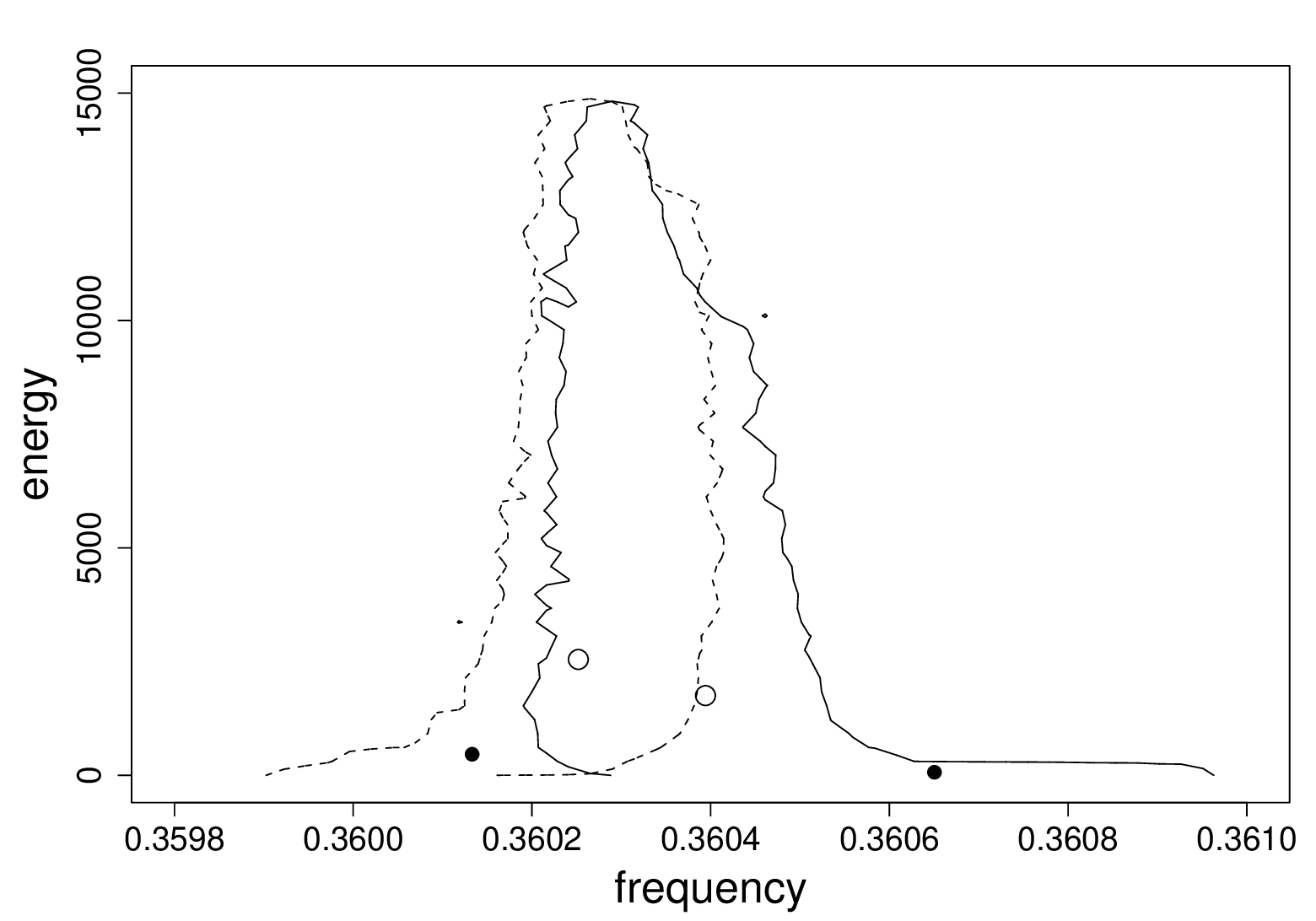}
\end{tabular}
\caption{95\% posterior confidence contours of energy and
frequency for the six sinusoids considered. The empty circles show
the median value of the posterior peak, and the filled circles
show the true value.} \label{fig_6}
\end{figure}

Although spectral estimation is an important topic, for LISA
analysis a more pertinent issue is the identification of orbital
parameters for the compact binary systems generating the signals.
In our example, we can therefore consider the joint posterior of
the frequency $f_i$ and the energy $E=N(A_i^2+B_i^2)/2$ for the
six sinusoids in our three regions (Fig.~\ref{fig_6}). Unlike in
Fig.~\ref{fig_5}, the ordinate now displays an energy rather than
an energy density, and the contour is at the 95\% credible level.

\begin{figure}[hbt]
\centering
\includegraphics[width=16cm]{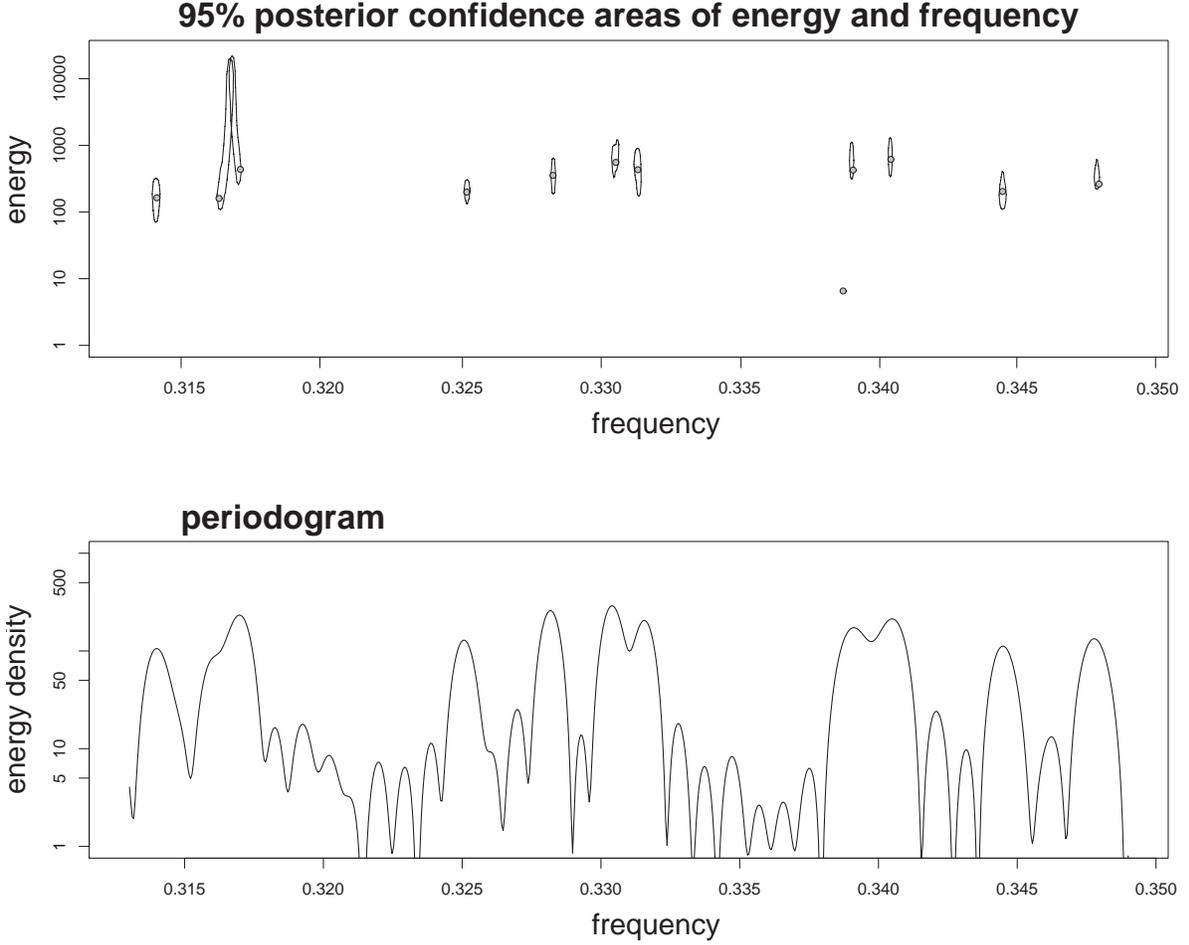}
\caption{Top: A section of the (energy, frequency) joint posterior
probability density containing eleven 95\% posterior confidence
areas. The bottom plot shows the classical periodogram for the
same frequency range.  Twelve sinusoids are actually present in
this region, indicated by gray dots and one sinusoid, at
$f=0.338\,720\,9$, was too weak to be identified by our method
($\gamma=0.11$).} \label{fig_7}
\end{figure}

To better point out the strength of this Bayesian approach, in
Fig.~\ref{fig_7} we display 95\% confidence areas for a wider
frequency range, including more sinusoids. As stated in
\cite{Bretthorst88}, ``if the signal one is analyzing is a simple
harmonic frequency plus noise, then the maximum of the periodogram
will be the best estimate of the frequency that we can make in the
absence of additional prior information about it''.  This is
demonstrated in Fig.~\ref{fig_7} showing the concordance in
frequency estimates for the well-separated sinusoids from both the
joint posterior and periodogram plots. However, the periodogram
peaks are significantly wider than the 95\% areas, and are clearly
sub-optimal for closely spaced sinusoids. In fact the posterior
probability density for frequency is the exponent of the ratio of
the periodogram $C(f)$ and the noise variance $\sigma^2$
\cite{Bretthorst88}. The Bayesian approach therefore takes account
of the noise variance in the estimation process, which explains
why the confidence regions are significantly narrower than the
periodogram peaks.

This Bayesian method therefore shows great power when tackling
strong signals closely spaced in parameter space (in this case,
frequency). In addition it delivers confidence intervals for the
parameters (frequency and amplitude) and can take account of
relevant prior information when applied to LISA data.

\begin{figure}[hbt]
\centering
\includegraphics[width=16cm]{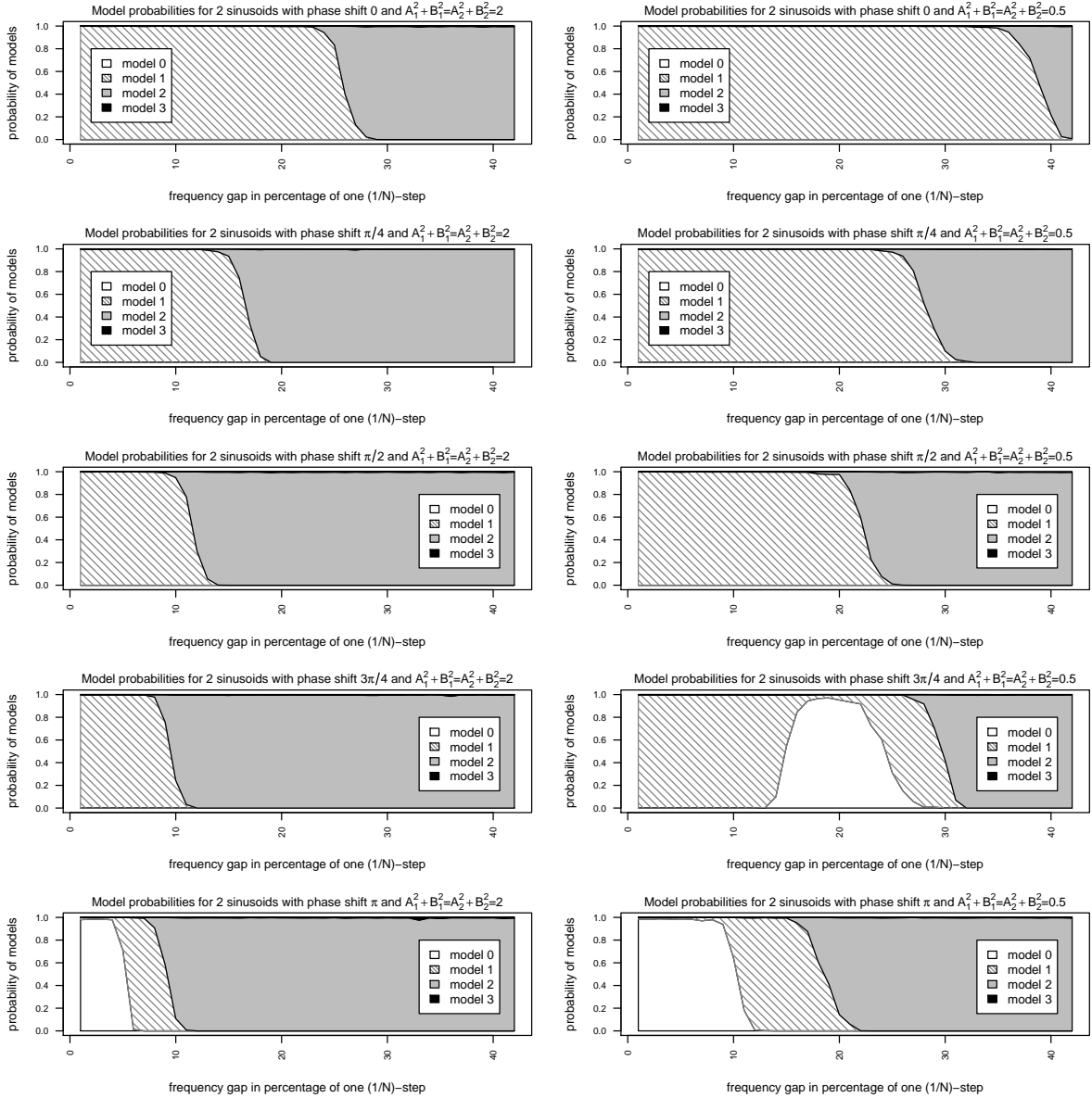}
\caption{Model probabilities for different pairs of
equal-amplitude sinusoids in gaussian noise ($\sigma=1$). Each
plot displays two sinusoids with varying frequency gaps expressed
as a percentage of one $(1/N)$-step (with $N=1\,000$). Each row
corresponds to a different phase shift between the sinusoids: from
top to bottom the phase shift is $0, \pi/4, \pi/2, 3\pi/4, \pi$.
The two columns display results with different signal strengths,
as given in the title of the plots.} \label{fig_8}
\end{figure}

We have seen that Fig.~\ref{fig_4}, \ref{fig_5}, and \ref{fig_6}
reveal strong interference between very closely spaced signal
pairs resulting in poor estimation of their parameters.
Nevertheless, the Bayesian approach succeeds in revealing even
these as separate sinusoids, at a level far beyond the ability of
a classical periodogram. To investigate this further, we ran a
series of simulations with two sinusoids gradually approaching
each other in frequency. The results of these simulations are
presented in Fig.~\ref{fig_8} which shows how the ability of the
method to separate the signals depends on their strengths, their
relative phase and on observing time. In our examples we use
$t\in\{0,\dots,N-1\}$. We ran 10 series of simulations for five
different phase shifts ($0, \pi/4, \pi/2, 3\pi/4, \pi$) between
the two equal-amplitude sinusoids, and for two different signal
strengths, with $\sigma=1$. During each set of simulations the
frequency gap between the two sinusoids was increased by 1\% of a
$(1/N)$-step. (Recall that $N=1\,000$ so that $1/N=0.001$).   The
left-hand column of Fig.~\ref{fig_8} displays the results of runs
in which the sinusoids have $A_1^2+B_1^2=A_2^2+B_2^2=2$. The
right-hand column  displays results for a quarter of this energy,
i.e., $A_1^2+B_1^2=A_2^2+B_2^2=0.5$.
A high probability for model 2 indicates that the two sinusoids
could be separated well. If model 1 has the highest probability
then the two sinusoids could only be identified as one signal,
whereas model 0 indicates that the two sinusoids effectively
annihilated each other with respect to the observation time and
the resulting signal is too weak to be identified against the
background noise. This last effect can only be observed for phase
shifts close to $\pi$ since the amplitudes then have opposite
signs and for a frequency gap of around zero almost no resultant
amplitude is developed within the observation time considered.

It is obvious that stronger sinusoids are easier to detect and
separate, so model 2 is favored with much smaller frequency gaps
in the left-hand column of Fig.~\ref{fig_8} than in the right-hand
column. Furthermore, in both the left and right-hand columns the
separation ability is worse for small phase shifts and generally
improves as the phase shift increases, but this again depends on
the noise level. The most striking effect can be seen in the right
column at a phase shift of $3 \pi/4$. For a small frequency gap
the sinusoids can be detected but not separated, then from 14\% of
a $(1/N)$-step model 0 starts to be favored. One might expect an
increase in the probability of of model 2 at this stage, but in
fact their frequency and phase separation mean the pair almost
annihilate each other over the time period considered, so that the
resultant cannot be seen against the noisy background. With
further separation the effect is reduced, and for larger frequency
gaps (30\% of a $(1/N)$-step) model 2 does indeed become dominant.
This phenomenon can also be seen in the periodogram revealing a
single peak slowly decreasing in height from 1\% to about 19\% of
a $(1/N)$-step. Then the signal has the same height as the
surrounding noise and is slowly increasing beyond 20\% of a
$(1/N)$-step. Note that for a different observing period, say  $t
\in \{t_{\rm start},\dots,t_{\rm start}+N-1\}$, the results would
have been different. However even in the worst case the detection
can be achieved within 40\% of a $(1/N)$-step for the weaker
sinusoids and within 26\% for the stronger sinusoids.

\section{Discussion}
\label{disc} In this paper we have presented a Bayesian approach
to identifying a large number of unknown periodic signals in a set
of noisy data. Our reversible jump Markov chain Monte Carlo method
can be used to estimate the number of signals present in the data,
their parameters, and the noise level. This method compares
favorably with classical spectral techniques. Our approach allows
for simultaneous detection and parameter estimation, and does not
require a stopping criterion for determining the number of
signals.

Although the parameters of even strong components are not well
determined when they are sufficiently close together in frequency,
we still obtain useful confidence intervals. Importantly, the
noise level is itself a parameter in the overall fit so that the
energy present in the data is automatically allocated to either
signal or noise. The large uncertainties in amplitudes we obtained
for the closely-spaced components in our example of 100 sinusoids
will, for a practical problem, be lessened by a sensible choice of
priors. A prior similar to the $g$-prior, using a diagonal matrix
and a hyper-parameter $g$ as a multiplicative factor, might be a
good choice. For the real LISA data we would use priors for the
signal amplitudes that would reflect our astrophysical knowledge.

The motivation for our research is to address the difficulty that
LISA will ultimately encounter in what is loosely called the
\emph{confusion problem}. LISA may see as many as 100\,000 signals
from binary systems in the 1\,mHz to 5\,mHz band. We therefore
view our work as a powerful new technique for identifying and
characterizing these signals in the LISA data stream.

The work presented here is of course a highly simplified toy
problem: we are neglecting signal modulation due to LISA's orbit
and beam pattern and are not considering an appropriate data model
for LISA. In addition, the signals from compact binary systems
differ from simple sinusoids. However, we believe that it
demonstrates the applicability of the approach to LISA data
analysis, and the next step is to deal with these more complicated
signals and to develop a realistic strategy for applying our MCMC
methods. We do not claim that this will be a trivial extension; in
fact, we acknowledge the complexity of the situation. However, we
believe that MCMC methods, like those presented here, do give a
realistic strategy for identifying and characterizing the large
number of signals, of all types, that will exist in LISA data.

Besides LISA, the methods discussed here are likely to be useful
in other fields of study where the data contain an unknown number
of periodic signals. One timely area of astrophysics is in the
field of asteroseismology, where attempts are made to measure
vibrational modes of a star; see for example Handler et al.\
\cite{handler04}. We would expect that our method could be
directly applied to asteroseismic data on stellar oscillations.
Similarly, spectroscopic nuclear magnetic resonance studies are
often concerned with spectral decomposition into an unknown number
of components, dependent on the composition of the sample
\cite{Levitt01}, and we believe the parameter estimation technique
we present could be usefully applied here too.

\begin{acknowledgments}
This work was supported by National Science Foundation grants
PHY-0071327 and PHY-0244357, the Royal Society of New Zealand
Marsden fund award UOA204, Universities UK, and the University of
Glasgow.
\end{acknowledgments}

%%%%%%%%%%%%%%%%%%%%%%%%%%%%%%%%%%%%%%%%%%%%%%%%%%%%%%%%%%%%%%%%%%%%%%%%
%                               BIBLIOGRAPHY
%%%%%%%%%%%%%%%%%%%%%%%%%%%%%%%%%%%%%%%%%%%%%%%%%%%%%%%%%%%%%%%%%%%%%%%%

\end{document}